\documentclass[3p,sort&compress]{elsarticle}

\usepackage{lineno}
\modulolinenumbers[5]
\usepackage{subcaption}

\journal{Journal of Non-Newtonian Fluid Mechanics}

\usepackage{xcolor}
\usepackage{graphicx} %package to manage images
\usepackage{float}
\usepackage{multirow}
\usepackage{hyperref}
\usepackage{color}
\usepackage{siunitx}

\usepackage{amsmath}
\usepackage{amssymb}

\usepackage{algorithm}
\usepackage[noend]{algpseudocode}

\usepackage{upgreek}

\makeatletter
\def\BState{\State\hskip-\ALG@thistlm}
\makeatother

\begin{document}

\begin{frontmatter}

%\title{Bifurcating flow of a yield-stress fluid: Experimental characterization of the solid-fluid transition}
\title{Atypical plug formation in internal elastoviscoplastic fluid flows over non-smooth topologies 
%over non-smooth topologies %{\color{red}Miguel: should the title be more specific on the geometry? Maybe sudden rectangular opening? or cavity and aperture}
}
\author[UBCCHBE]{Miguel E. Villalba}

\author[UBCMATH]{Masoud Daneshi\corref{mycorrespondingauthor}}
\cortext[mycorrespondingauthor]{Corresponding author: masoud.daneshi@ubc.ca}

 \author[STRATH]{Emad Chaparian}
 \author[UBCCHBE]{D. Mark Martinez}

\address[UBCCHBE]{Department of Chemical and Biological Engineering, University of British Columbia, Vancouver, BC, Canada, V6T 1Z4}
\address[UBCMATH]{Department of Mathematics, University of British Columbia, Vancouver, BC, Canada, V6T 1Z2}
\address[STRATH]{James Weir Fluid Laboratory, Department of Mechanical \& Aerospace Engineering, University of Strathclyde, Glasgow, United Kingdom}

\begin{abstract}
{An experimental and computational investigation of the internal flow of elastoviscoplastic fluids over non-smooth topologies {\color{black}is presented in two complimentary studies}. %Our primary motivation is to gain insight into the behaviour of suspensions in filtration, where fouling can happen and refers to the formation of stagnant plug zones in sudden openings or apertures. %We first 
{\color{black}In the first study, we} visualize the {\color{black}creeping} %2D 
flow of {\color{black}a} Carbopol gel %s
%, a model fluid of colloidal suspension, 
over a cavity embedded in a thin slot using Optical Coherence Tomography (OCT) {\color{black}and confocal microscopy}. %We also visualize the flow field of Carbopol gels over an aperture embedded in a thin slot using confocal microscopy. 
We measure the size and shape of the plug as a function of Bingham and Weissenberg numbers. An asymmetry in the plug shape is observed which is also evident in our second study---numerical simulations using adaptive finite element method based upon an augmented Lagrangian scheme. We quantify the asymmetry and present the results as a function of the product of the Weissenberg and Bingham numbers which collapse onto a single curve for each of these geometries. These findings underscore the theoretical underpinnings of the synergy between elasticity and plasticity of these complex fluids.}
\end{abstract}


\end{frontmatter}

%\linenumbers

\section{Introduction}

%In this work, we examine the fouling of a cross-flow filtration aperture found in a number of natural and industrial settings~\citep{BHAVE2014149}. 
{\color{black}In this work, we examine the formation of a plug created by the flow of a yield stress fluid in small cavity, i.e. a geometry found in a number of natural and industrial settings~\citep{BHAVE2014149}.}
The primary motivation for this work stems from an industrial application, namely pressure screens which are commonly found in the pulp and paper industry. Here, fouling of the screen {\color{black}apertures} %surface 
reduces both the capacity and efficiency of the screening process. 
%Although jamming events for dry granular materials or colloidal suspensions are well-investigated, we find that for non-Brownian {\color{black}fibre or rod-like suspensions} %suspension flows these events are relatively unexplored. The essential difference between these two bodies of literature is that suspension flows can jam under dilute conditions. 
{\color{black}Key to unravelling this is the understanding of the role of the rheology of the suspension on the formation of a local plug. A quite general characteristic of the rheology of suspensions with high enough solid loadings or with sufficiently strong inter-particle interactions is the yield stress \cite{Ovarlez2013,Derakhshandeh2012,Emady2013}. }
%Key to unravelling this is the understanding of theinteraction between the rheology of the suspension and the resulting frictional forces leading to a local jam.

%%In principle, 
%{\color{black}We note that the rheology of} suspensions with high enough solid loadings or {\color{black}with} sufficiently strong inter-particle interactions can {\color{black}adopt a} %manifest 
%yield stress due to the network structures formed by {\color{black}either} mechanical and chemical {\color{black}interactions} %forces
% \cite{Ovarlez2013,Derakhshandeh2012,Emady2013}. 
%It is well known that y
{\color{black}Y}ield-stress fluids are characterized by a transition from solid-like to fluid-like behaviour above a threshold, $\tau_y$---the yield stress. This implies a co-existence of a liquid and solid phase, which may cause the material to {\color{black}plug small apertures in a pressure driven flow }%clog up the aperture or conduit and interfere with the flow if the driving pressure gradient is too small 
\cite{ROUSTAEI2013109,Daneshi2020}. 
{\color{black}Beside the yield stress, these materials exhibit %more 
complex properties such as} elasticity and thixotropy \cite{OVARLEZ201368,Putz2009,Balmforth2014} {\color{black}which} further complicate their flow features and {\color{black}our understanding of} solid-liquid transition {\color{black}\cite{Coussot2014,zare2021effects,daneshi2023growth}}. 
In this study, we focus on {\color{black}characterising} the stagnant regions forming in the apertures and their link to {\color{black}the rheology of the material}. To do so, we {\color{black}simplify the cross-flow geometry and reduce this problem to} an idealized confined flow of a yield-stress fluid  over a cavity, i.e.~a sudden expansion-contraction {\color{black}in a thin slot}.  

A large body of numerical works have studied the flow behaviour of yield-stress fluids in expansions and contractions by exploiting the constitutive rheological models such as the Bingham and Herschel-Bulkley models  \cite{ROUSTAEI2013109,Abdali1992,Mitsoulis1993,MITSOULIS2001173,VIGNEAUX201838,hewitt2016,balmforth2017}. These studies predicted the plug regions to be symmetrical {\color{black} with respect to the geometry of the expansion-contraction}
 in the absence of inertia and showed how the growth and shrinkage of these regions depend on the yield stress of the material. 
However, recent experimental investigations demonstrated anomalous complexities including the asymmetry of the yielding surfaces in the flows developed over the cavity \cite{Jay2001,Alexandrou2001,DeSouzaMendes2007,Luu2015,Varges2020}. In these cases, stagnant zones form in the corners of the expansions and contractions. These stagnant regions develop into an internal asymmetric plug over the cavity and are separated from the yielded regions by a discernible interface {\color{black}\cite{Luu2015,Philippe2017,Varges2020}}. 
%They speculated that t
{\color{black}T}hese complexities %may 
arise from the complex rheology {\color{black}of the practical yield-stress fluids} in terms of their elasticity/viscoelasticty or thixotropy which are not captured by conventional (i.e.~``simple") viscoplastic models. {\color{black}In this context, the flow over a cavity, the problem of interest in this study, can prove to be a fertile playground to better understand the effect of complex non-ideal rheology of practical yield-stress fluids on the flow hydrodynamic features.} 

Asymmetry is not limited to the flow of yield-stress fluids over cavities and has also been reported previously in various configurations. 
The flows of Carbopol gels around obstacles have been a subject of %meticulous 
experimental investigations for both unconfined  \cite{Tokpavi2009,Ahonguio2014,TOKPAVI200935,JOSSIC201314} and confined settings \cite{Daneshi2020}. Interestingly, a fore-aft asymmetry was observed between the upstream plug and the downstream plug behind the obstacle \cite{Tokpavi2009} regardless of surface roughness \cite{Ahonguio2014} and the shape or number of obstacles \cite{daneshi2019characterising,Daneshi2020}. Daneshi et al.~\cite{Daneshi2020} showed that by increasing the yield stress or {\color{black}by slowing} % down 
the flow%(i.e. increasing the Bingham number)
, the extent of asymmetry intensifies. 
In general, the key finding of all these studies is that flow asymmetry is robust regardless of geometry or confinement even in the creeping flow regime. The common conjecture in these studies is that the complex rheology effects, such as elasticity and thixotropy, lead to the asymmetrical flow. However, characterization of this asymmetry and its link to the rheological parameters of the material remain largely unexplored. 
 
Recent %pioneering 
numerical works have also demonstrated asymmetry in the flow of yield-stress fluids by using novel elastoviscoplastic models \cite{cheddadi,Fraggedakis,izbassarov2018computational,chaparian2019porous,chaparian2019adaptive}.  
The elastoviscoplastic models were first introduced by Saramito \cite{SARAMITO2007,saramito2009new} and were obtained from combining the Bingham/Herschel-Bulkley model \cite{bingham1922} and the Oldroyd model \cite{oldroyd1950} which includes the effect of yielding and elastic behaviour of the material, respectively. %Hence, t
{\color{black}T}hese elastoviscoplastic models can predict elastic creep and recoil below the yielding point as well as the viscoelastic relaxation in the liquid regime.
Crucially, Cheddadi et al.~\cite{cheddadi} compared the viscoplastic (Bingham model), viscoelastic (Oldroyd-B model) and elastoviscoplastic flows around a circular obstacle. They reproduced similar flow asymmetries observed in the experimental counterpart of their work for a practical yield-stess fluid and demonstrated that these flow asymmetries are only present in the elastoviscoplastic flows in the case of low Weissenberg numbers \cite{cheddadi}. These findings challenge the common conjecture that elasticity is the only cause for asymmetry and suggest that a combination of elasticty and plasticity of the fluid comes to play a role. More recently, Chaparian \& Tammisola \cite{chaparian2019adaptive} quantified the asymmetry in the context of elastoviscoplastic fluid flows through wavy channels and shed more light on the characterization of the hydrodynamic complexities of these type of flows in terms of fluid rheology. In this work, we aim to take this analysis further more systematically from both the experimental and numerical perspective. 

Here, we investigate the internal flow of yield-stress fluids around {\color{black}a} %the 
cavity/aperture, {\color{black}in a regime where plug formation is possible.} %where there is a possibility of plug formation. 
Our objective is to improve the understanding of the plug phenomenology and the asymmetry in the flows and yield surfaces. 
In particular, we attempt to formulate the extent of the asymmetry in terms of the rheological properties of the working fluid and the flow parameters. 
An experimental flow visualization technique and a previously developed numerical method \cite{chaparian2019adaptive} have been adapted to study the 2D flow of Carbopol gels over a cavity. %Principally, the e
{\color{black}E}ffects of different rheological parameters on the formation and shape of the unyielded surfaces are examined %. Using a thorough analysis of experimental and numerical data, we try to find 
{\color{black}and} a link between the governing parameters of the problem and the extent of asymmetry {\color{black}proposed}. 
In addition, we extend our work to the flow in a more complex geometry, namely a 3D flow confined in a Hele-Shaw cell aperture with a step. The objective is to examine whether our findings in the quasi 2D flow hold valid in a 3D cross-flow filtration geometry as well. 

The paper is outlined as follows. Section \ref{sec:exp} describes the fluids used in our experiment, their rheological properties, experimental set-ups, and the methods implemented to analyze the experimental data. Section \ref{sec:results} presents the experimental results and the numerical simulations of the 2D flow over a cavity. Finally, concluding remarks are made in the last section.

\section{Experimental Design \& Materials}\label{sec:exp}

\subsection{Material and rheology}
\label{Sec:Mat}

In this work, we use Carbopol gel which has been widely known as a model yield-stress fluid due to its optical transparency and neglegible thixotropic behaviour \cite{daneshi2019characterising}. This material has been widely used in many visualization experiments and rheological studies \cite{OVARLEZ201368,COUSSOT201431}. Carbopol gels of two different concentrations $0.060 \: (wt/wt\%)$ and $0.075 \: (wt/wt\%)$ were %carefully 
prepared according to the procedure explained in our previous works \cite{Daneshi2020,daneshi2019characterising}. The rheological curves and parameters of the fluids were measured using a rotational rheometer (MCR501, Anton Paar) with an angular resolution of $0.01 \: mrad$ and a torque resolution of $0.1 \: nNm$. A parallel plate geometry with the diameter of $50 \: mm$ was used for the rheological measurements with a ramp rate of $0.05 \: Pa/s$. To remove any effect of wall slip, %a piece of 
sandpaper with an average roughness of $35\;\mu m$ was glued onto the plates. 

\begin{figure}[h]
    \centering
    \includegraphics[width=12 cm]{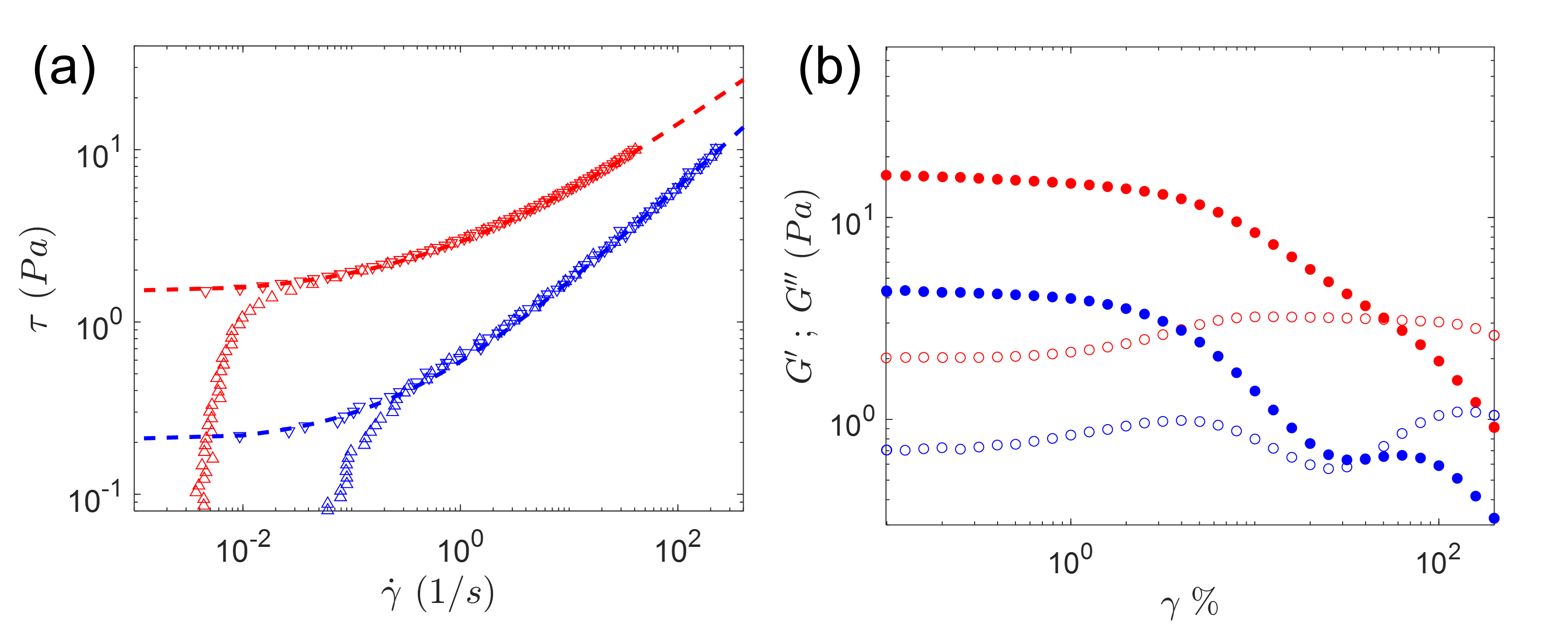}
    \caption{{\color{black}Rheology of Carbopol gels used in our experiments:} Flow curves for two different Carbopol solutions measured by a stress controlled ramp-up and ramp-down test with a ramp rate of $0.05 \ Pa/s$ {\color{black}shown in panel (a). Blue symbols represent the $c=0.060\%$ Carbopol, while the red symbols represent the $c=0.075\%$ Carbopol. The upward and downward triangle denote the ramp-up and ramp-down data, respectively}. The dashed lines show the Herschel-Bulkley fits.  {\color{black} Measurements of the
shear storage (closed symbol) and loss moduli (open symbol), G' and G", from small amplitude oscillatory rheometry with a frequency of $1\;Hz$ are shown in panel (b)}.
}
    \label{fig:flowcurve}
\end{figure}

A shear stress ramp-up and ramp-down tests were performed to measure the flow curves for the two Carbopol gels used in this work. The results are shown in Fig.~\ref{fig:flowcurve}a. 
The rheological behaviour of Carbopol gels during the ramp-down test was well characterised by the Herschel-Bulkley constitutive law. 
The ramp-up curve and ramp-down curve overlap above the yielding point which is suggestive of the non-thixotropic behaviour %of this material 
over this region. 
The elastic response of the material manifests itself as finite deformation at the start of the ramp-up curve and as elastic recoil at the end of the ramp-down curve {\color{black}\cite{Balmforth2014,Lectures_Ovarlez,Carbopol_divoux}}. Note that the negative shear rates obtained at the end of ramp-down curve are absent in Fig.~\ref{fig:flowcurve} due to the logarithmic scale. 
The Herschel-Bulkley fits of the yield stress, $\tau_y$, consistency, $K$, and powerlaw index, $n$, are given in Table \ref{tab:1}. Moreover, we measured the shear storage and loss moduli, $G'$ and $G''$, from a small amplitude oscillatory rheometry test {\color{black}(see Fig.~\ref{fig:flowcurve}b)} at a frequency of $\omega =1 \ Hz$ and a strain amplitude of $\gamma = 0.01 \%$. These parameters are also reported in Table \ref{tab:1}. 

In addition to Carbopol gels, we conducted a small number of experiments with two other fluids to benchmark our observations. A polyethelene oxide solution (PEO) of $c = 0.75 \: (wt/wt\%)$ was prepared as a prototypical viscoelastic fluid with no yield stress. A glycerol-water solution of $c = 31 \: (wt/wt\%)$ was prepared as a prototypical viscous Newtonian fluid.

%% new table with other fluids 
\begin{table}[]
\centering
\caption{\color{black}List of the fluids used in our experiments and their rheological properties. Note that the Herschel-Bulkley parameters were obtained from a regression with an R-squared value greater than $0.99$. The shear storage and loss moduli ($G^{\prime}$ and $G^{\prime \prime}$) measurements taken from small amplitude oscillatory rheometry at a frequency of $1\;Hz$ and a strain amplitude of $\gamma = 1\%$, within the linear regime. Note that the characteristic relaxation time of PEO has been measured from a frequency sweep test in \citep{hormozi2011} and reported to be in the order of a few seconds. %$3\pm 0.5\;s$. 
It was estimated as the the inverse of the frequency at the intersection of the elastic and loss moduli when plotted versus frequency. 
%(below which we confirmed that the two moduli where independent of γ).
%
%List of the fluids used in our experiments and their rheological properties. Note that the parameters were obtained from the regression using a Hershcel-Bulckely model. The goodness of fit R-squared was found to be greater than 0.99.  
} \label{tab:1}
\small
\begin{tabular}{ccccccc}
\hline  
Fluid & \text{\begin{tabular}[c]{@{}c@{}}Concentration\\ $(wt/wt\%)$\end{tabular}} & \textbf{\begin{tabular}[c]{@{}c@{}}$\tau_y$\\ $(Pa)$\end{tabular}} & \textbf{\begin{tabular}[c]{@{}c@{}}$K$\\ $(Pa s^n)$\end{tabular}} & \textbf{$n$} & \textbf{\begin{tabular}[c]{@{}c@{}}$G^{\prime}$\\ $(Pa)$\end{tabular}} & \textbf{\begin{tabular}[c]{@{}c@{}}$G^{\prime \prime}$\\ $(Pa)$\end{tabular}} \\ \hline 
\multirow{2}{*}{Carbopol} & 0.060 & 0.20 & 0.39 & 0.57 & 4.2 & 0.7 \\  
 & 0.075 & 1.47 & 1.53 & 0.46 & 16 & 2.0 \\ \hline 
PEO & 0.75 & 0 & 1.02 & 0.49 & {\color{black}0.85} & {\color{black}0.24} \\ \hline
\multicolumn{1}{l}{Glycerol-Solution} & 31 & 0 & {\color{black}0.002}  & 1 & - & - \\ \hline
\end{tabular}
\end{table}

\subsection{Experimental Setup and methodology}
\label{Sec:Exp_setup}
In this work, we study two different types of flows: a 2D flow over a cavity and a 3D flow in a thin slot {\color{black}past }%over
 an aperture (see Fig.~\ref{fig:setup}). These topologies are embedded in two channels where the far field flow is unidirectional.

Direct visualization of the flow using state-of-the-art imaging devices was exploited to characterize the velocity fields and yielding surfaces developing around the cavity and aperture \cite{Daneshi2020,daneshi2019characterising}.
For each of these geometries, the details of the experimental setup including the channel geometry and its fabrication as well as the visualization techniques employed to monitor the flow are explained in the following subsections. 

\subsubsection{The 2D flow over a cavity set-up}\label{subsec:2dexp}

\begin{figure}[h]
    \centering
    \includegraphics[width=11 cm]{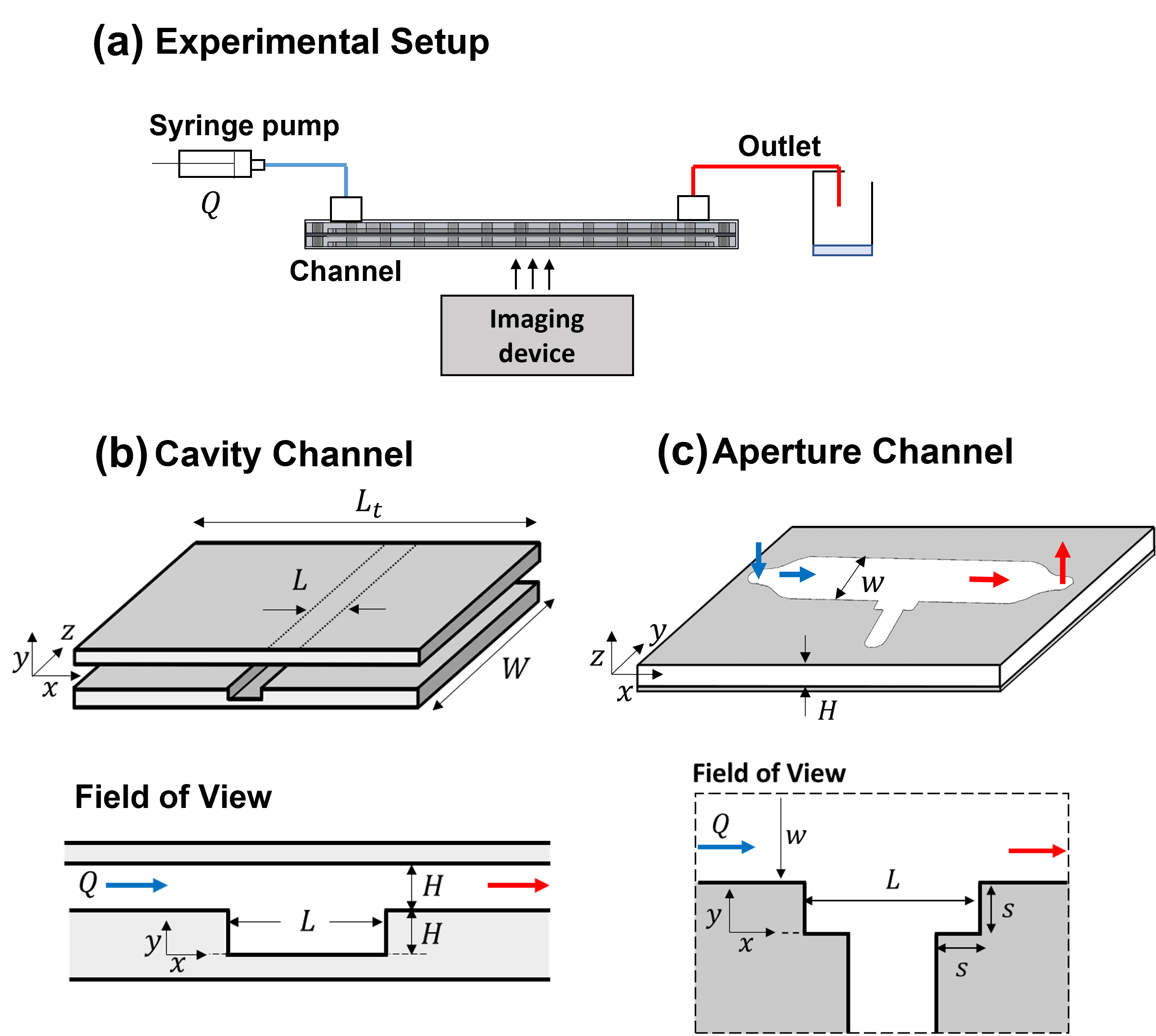}
    \caption{A schematic diagram of the experimental set-up is shown in panel (a). The flow is generated by a syringe pump and monitored in a thin slot. The thin slot includes either a cross-wise cavity or a planar aperture. The schematic diagrams of the 2D flow in a cavity and the 3D flow through an aperture are depicted in panels (b) and (c), respectively. We used optical coherence tomography and florescent confocal microscopy to image the flow (in $x-y$ plane) around the cavity and planar flow (i.e.~$x-y$ plane) around the aperture, respectively. The corresponding fields of view for these geometries are also shown in these two panels. Note that blue arrows show the incoming feed flow, while red arrows show the direction of the outgoing flow. For the cavity, $L=4 \: mm$ while for the aperture $L=12 \: mm$.
}
    \label{fig:setup}
\end{figure}

The cavity in the channel (see Fig.~\ref{fig:setup}b) consists of a thin ($1 \: mm$ thick) aluminium plate enclosed by two clear ultra-scratch-resistant cast acrylic plates ($6.3 \: mm$ thick). The aluminium plate separates the acrylic plates creating a flow cell of thickness $H=1 \: mm$. The channel has a width $W= 64 \: mm$ and length $L_t=105 \: mm $. This channel has one outlet which is open to atmosphere at the same height as the inlet. A cavity was machined in the center of one of the acrylic plates (see Fig.~\ref{fig:setup}b) by means of computer numerical control (CNC) milling machine. The cavity, which spans the whole width of the channel, has a depth of $H =1 \: mm$ and length of $L=4 \: mm$. The surface of the cavity was roughened using a sandpaper to inhibit the possibility of  wall-slip.
The distance from the entrance of the channel to the cavity is long enough to ensure fully-developed flow at the upstream of the cavity. {\color{black}Note that the surfaces of the bottom and top plates are not treated and we expect that the gel shows wall-slip behaviour. As reported in Ref.~\citep{Daneshi2020}, this might change the velocity field at the upstream slightly, but does not hinder the flow development.}

For this experiment, the flow of Carbopol gel is supplied to the cell via a syringe pump in the range of $Q = 0.01-0.3 \: ml/min$. Table \ref{tab:2} summarizes the experimental conditions including the flow rates and the corresponding range of mean velocities $U_c$. 
Before imaging, the channel was flushed with ethanol and then distilled water. 
To start each experiment, we filled the channel with Carbopol gel at a fixed flow rate $Q$. Then, after giving the flow enough time to reach steady state, the flow visualization was commenced. 

For imaging the flow, we used white polystyrene beads (Magspheres), with the mean diameter of $4 \;\mu m$ at a concentration of $0.005 (wt/wt\%)$, as the tracing particles. The particles were well mixed and the samples were subsequently sonicated to ensure homogeneity. We visualized the flow in the longitudinal centre-plane of the cell to measure the cross-slot streaklines and velocity fields developing over the cavity (see bottom panel of Fig.~\ref{fig:setup}b). Since the observation plane is in the centre of the channel, we expect that the effect of the side walls are negligible and the flow can be considered to be 2D.

Flow visualization was done using a {\color{black}Thorlabs TEL1300V2-BU}
 Optical Coherence Tomography (OCT). The OCT is a real-time imaging device that works based on inter-ferometry with a broad bandwidth light source. The OCT focuses a collimated beam of light with the center wave length of $1300 \: nm$ to the specimen by using a 5x-objective lens. The lateral scanning of the sample is performed by moving vertical scanning beam laterally through a two-galvo mirror system. The device captures cross-sectional images from the backscattered light coming from the tracing beads. The depth of focus of the vertical imaging is approximately $3.5 \: mm$ in a medium with the same refractive index as air, while the lateral field of view is adjusted to be $5 \: mm$. The spatial resolution of vertical imaging of the sample is around $3.5 \ \mu m$ while its lateral resolution is around $13 \ \mu m$. The images of flow were automatically recorded to a hard drive at a rate of $10 \: Hz$.
 %with micrometer-scale axial and transverse resolution and millisecond temporal resolution \cite{Daneshi2020}.  
 {\color{black}We refer the
reader to Ref.~\cite{daneshi2019characterising,OCT1} for more details on imaging by OCT and its application to flow visualization.}

\subsubsection{The 3D flow {\color{black}{past}} %over 
aperture set-up} \label{subsec:3dexp}

The aperture flow experiments were carried out in a narrow cell, {\color{black}{as depicted }}%which its schematic is shown 
in Fig.~\ref{fig:setup}c. 
The cell was constructed using a thin Teflon PTFE sheet which is enclosed by two clear ultra-scratch-resistant cast acrylic plates ($6.3 \: mm$ thick). The acrylic plates include grooves and o-rings for liquid sealing. The acrylic plates and spacer were installed in an aluminium frame. The channel has a width of $w = 16.38 \: mm$, length of $L_t=105 \: mm $ and height of $H=1 \: mm$. 
%The distance between the inlet and the aperture is long enough to ensure fully developed flow. 
%The channel geometry consists of a single aperture with a square step (see bottom panel of Fig.~\ref{fig:setup}c).
{\color{black} The channel's geometry features a square step aperture (see bottom panel of Fig.~\ref{fig:setup}c) with a cavity and a bottom slot, which closely resembles a screening aperture in design.}

A Teflon sheet was cut to the shape of the aperture by means of a computer controlled waterjet cutter. {\color{black}{The surface of the aperture was roughened using a sandpaper to inhibit the possibility of  wall-slip}}. 
 The details of the aperture geometry are shown in the lower panel of Fig.~\ref{fig:setup}c, where the span of the opening $L = 6 \: mm$ and the step size $s = 3 \:mm$.
Similar to the cavity of Fig.~\ref{fig:setup}b, the ratio of the depth ($s$) to the total length ($L$) of the aperture is 1/4. The channel is wide and thin enough such that the flow in the upstream region is unidirectional, while the flow {\color{black}{past}}
%over
the aperture is expected to be 3D.
%The aperture has a span of $L_s = 6 \: mm$, and the length of each of the adjacent square steps is $s = 3 \:mm$. Overall, the total length of the aperture is $L=12  \: mm$. Similar to the cavity of Fig \ref{fig:setup}b, the ratio of the depth and total length of the geometry is 1 to 4. The height of this channel is $H = 1 \: mm$. 

Before running each experiment, the channel was cleaned with distilled water and ethanol. Then, the channel was filled with the working fluid via a syringe pump at a  flow rate of $Q= 1 \: ml/min$. Subsequently, the working fluid was pumped into the cell at a lower constant flow rate for approximately $20\; mins$ before the flow visualization commenced. The range of flow rates in these experiments are reported in Table \ref{tab:2}. 

For imaging the flow, we used fluorescent beads (Magspheres), with a mean diameter of $2.9 \;\mu m$ seeded at a concentration of $0.005 \: (wt/wt\%)$, as the tracing particles. Flow visualization was conducted using a swept-field laser-scanning confocal microscope {\color{black}(Nikon Ti Eclipse)}
with a $4$X-objective lens. 
%Because the field of view of the lens $(2 mm \times mm)$ was too small to capture the entire region of interest, the full flow field over the region of interest was created by stitching together a series of images. 
The observation plane was fixed at the central horizontal $x-y$ plane, but its lateral position is controlled by a motorized stage. Since the field of view of the lens is smaller than the size of the aperture, the full flow field was created by stitching together the images of smaller areas. The spatial and temporal resolution of the device is $1.6 \;\mu m$ and $50 \; ms$, respectively. {\color{black}We do not give details of imaging and flow visualization using Confocal microscopy here, and refer the readers to Refs.~\cite{Daneshi2020,ConfocalI,ConfocalII} for more information.}   

\begin{table}[]
\centering
\caption{Experimental matrix: list of geometries, Carbopol gels and their rheological properties, flow parameters including the ranges of flow rates and average velocities in the upstream of the flow cell $U_c$, and ranges of governing dimensionless groups including $B$ and $Wi$. %The HB parameters were obtained from the best Herschel-Bulkley fit of the Carbopol solutions with R-squared better than 0:99.
} 
\label{tab:2}
\small
\begin{tabular}{cccccc}
\hline
Experiment & \text{\begin{tabular}[c]{@{}c@{}}Carbopol\\ $(wt/wt\%)$\end{tabular}} & \textbf{\begin{tabular}[c]{@{}c@{}}$Q$\\ $(ml/min)$\end{tabular}} & \textbf{\begin{tabular}[c]{@{}c@{}}$U_c$\\ $(mm/s)$\end{tabular}} & \textbf{$B$} & \textbf{\begin{tabular}[c]{@{}c@{}}$Wi$\\ $(10^{-3})$\end{tabular}}  \\ \hline 
\multirow{2}{*}{Cavity}  & 0.060    & 0.025-0.30  & 0.0065-0.13  & 9.04-1.64  & 0.27-5.4 \\  
 & 0.075  & 0.010-0.30  & 0.0026-0.08  & 14.4-3.19  & 0.08-2.4  \\ \hline
\multirow{2}{*}{Aperture} & 0.060  & 0.010-0.30  & 0.010-0.30  & 7.01-1.05  & 0.14-3.9  \\  & 0.075   & 0.010-0.30  & 0.010-0.30  & 7.71-1.67  & 0.10-2.9   \\ \hline
\end{tabular}
\end{table}

%Finally, Particle Image Velocimetry software (Dantec Dynamics) along with a MATLAB post-processing code were used to analyse the image-series and produce the velocity fields of the flow. The profiles of the yielding surfaces from both the raw images and the velocity contours were analyzed in MATLAB.

%1. PIV technique + Matlab code to generate velocity field
%2. how to determine the yielding surfaces:

\subsection{Flow and plug characterization}
\label{Sec:process_meth}

The incoming flow at the upstream of the cavity/aperture was studied before in a similar geometry as in Daneshi et al.{\color{black}~\cite{Daneshi2020,daneshi2020thesis}}. They showed that the velocity at the upstream of the channel is unidirectional and fully developed such that its cross-wise velocity profile matches the analytical one obtained from the Poiseuille flow of a Herschel-Bulkley fluid. Moreover, the fluid was pre-sheared before injection into the cell and along the tubes connecting the syringe to the channel at a higher shear rate than the nominal one in the channel. This might lead to the removal of the shear history of the fluid and minimization of any effects of possible thixotropic behaviour of the fluid.

In this work, we characterize the position of the static yield surfaces and the flow field developing around the cavity/aperture. The streaklines of flow of Carbopol over{\color{black}{/past}} the cavity/aperture were produced by simply taking the average of the series of the images of flow. The corresponding velocity fields were obtained by using a particle image velocimetry (PIV) technique implemented in a commercial analysis package (Lavision Davis 8.0) {\color{black}{\cite{PIVmethod}}}. A MATLAB code was used to post-process the PIV data and generate the velocity fields and contours.

% ?{\color{red}?????????Finally, the plug profiles from the raw images were tracked by applying edge detection and light intensity detection algorithms developed in MATLAB. The plug profiles from the velocity fields were identified at the points of low velocity.}

\begin{figure}[h]
    \centering
    \includegraphics[width=13 cm]{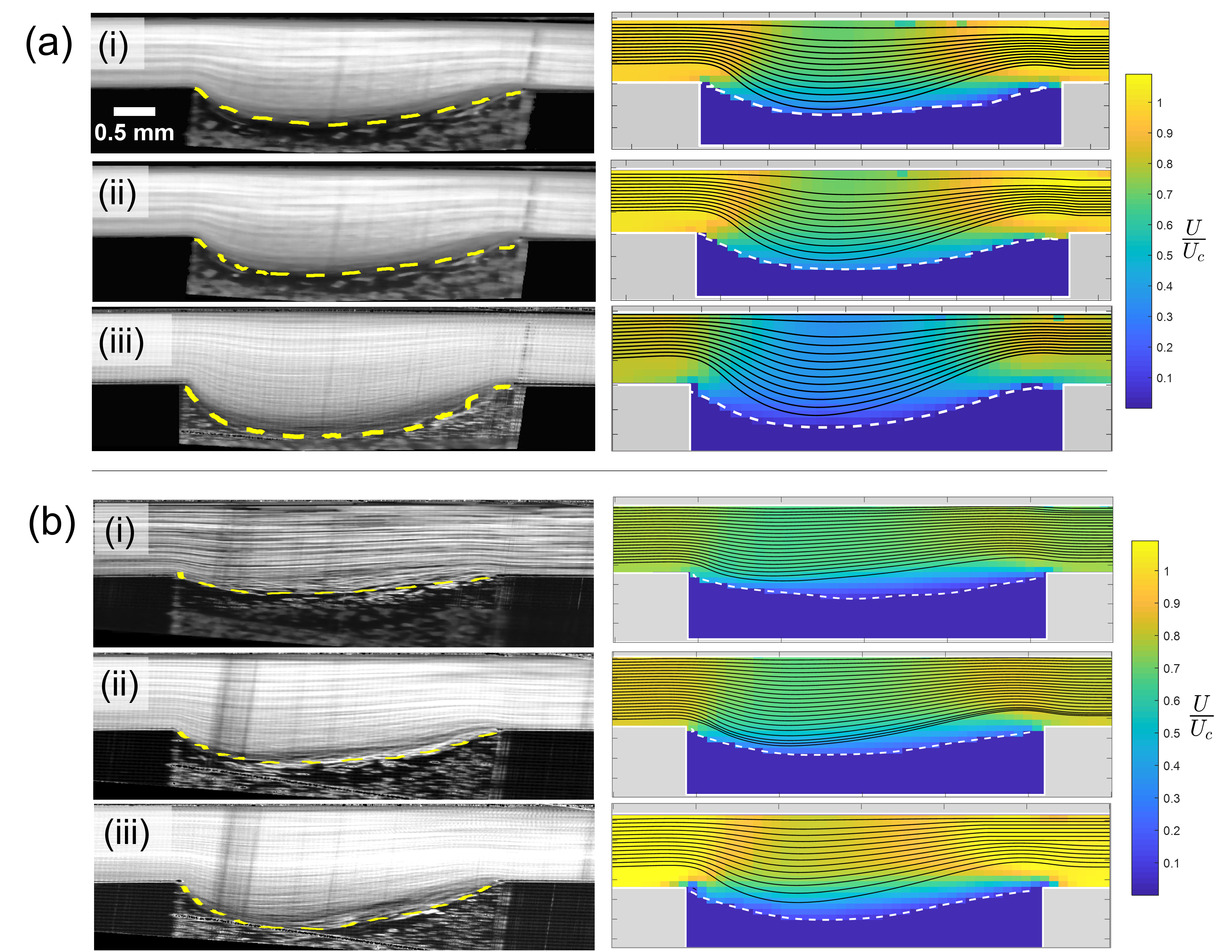}
    \caption{Representative experimental results for the flow of (a) $0.060\%\;(wt/wt)$ and (b) $0.075\%\;(wt/wt)$ Carbopol gels over a cavity. The images on the left panels show the streaklines of the flow while those on the right panels present the dimensionless speed contours, $U/U_c$.  For the $0.060\%\;(wt/wt)$ Carbopol, (i) $U_c = 0.006 \ mm/s$, $B = 9$, (ii) $U_c = 0.03 \ mm/s$, $B = 4.1$ and (iii) $U_c = 0.05 \ mm/s$, $B = 2.76$. For the $0.075\%\;(wt/wt)$ Carbopol, (i) $U_c = 0.003 \ mm/s$, $B = 14$, (ii) $U_c = 0.03 \ mm/s$, $B = 5$ and (iii) $U_c = 0.05 \ mm/s$, $B = 3.6$. The yellow and white dashed lines highlights the yield surfaces. %The color bar represent normalized velocity values.
    }    
%    
%    
%    Representative results for the flow of (a) $0.060\%\;(wt/wt)$ and (b) $0.075\%\;(wt/wt)$ Carbopol gels over a cavity. The images on the left side of the panel show the streaklines of the flow while the right ones show the dimensionless speed contours, $U/U_c$. For the $0.06\%\;(wt/wt)$, (i) $U_c = 0.006 \ mm/s$, $B = 9$, (ii) $U_c = 0.03 \ mm/s$, $B = 4.1$ and (iii) $U_c = 0.05 \ mm/s$, $B = 2.76$. For the $0.075\%\;(wt/wt)$, (i) $U_c = 0.003 \ mm/s$, $B = 14$, (ii) $U_c = 0.03 \ mm/s$, $B = 5$ and (iii) $U_c = 0.05 \ mm/s$, $B = 3.6$. The yellow and white dashed lines highlights the yielding surfaces. %The color bar represent normalized velocity values.
%    }
    \label{fig:plug2D}
\end{figure}

The profiles of static plugs (yield surfaces) were obtained using two methods which are similar to those developed in \cite{Daneshi2020}. In the first method, the yield surfaces were extracted directly from the streakline images. By means of a light intensity thresholding algorithm, the stationary tracer particles (seen as bright points) were separated from the moving tracers (seen as gray streaks). In the second method, the yielding surfaces were computed from a noise-based threshold in the velocimetry measurements. The two methods show good agreement with each other (see Fig.~\ref{fig:plug2D}). Finally, using the plug profile data and velocity measurements, we determined the size and shape of the unyielded regions.

Sample images of the flow of Carbopol gels over a cavity are shown in Fig.~\ref{fig:plug2D}a and Fig.~\ref{fig:plug2D}b for the $0.060 \%\;(wt/wt)$ and $0.075 \%\;(wt/wt)$ gels, respectively. The figures on the left depict the streaklines of the flow obtained from averaging the flow images. The images on the right depict the corresponding flow fields with the normalized velocity contours. As is clear from these figures, a plugged region forms inside the cavity, where the bright tracer particles become immobilized and the velocity approaches zero. Outside this region there is flow which is identified by the movement of the tracer particles. The static yield surfaces which separates the plugged region and flow region are highlighted in these figures by a dashed line.

%The plug profile or yielding surface was determined for both the raw images and the velocity fields. For For the streakline case, the interface was extracted directly from the averaged image streaklines by similar image analysis techniques as those used in \cite{Daneshi2020}. The interface in the velocity fields were obtained by identifying zones where the normalized velocity dropped substantially to values below the order of $\mathcal{O}(10^{-5})$. Although the position of the interfaces obtained from the two methods are not exactly identical, both exhibit the same morphology and behaviours. For clarity, we will use the velocity measurements to further characterize the morphology and phenomenology of the plugs.
\section{Results and Discussions}\label{sec:results}

%we present the experimental and numerical results on cavity
%also experiments on aperture
%numerics we implement an EVP constitutive law and for experiments we used carbopol gels

%The results of the visualization and numerical studies of this paper are presented below. Firstly, we report the visualization of the flow of YSF in a quasi 2D cavity. Secondly, we present the numerical simulation of the flow of elasto-viscoplastic fluid in this geometry. Lastly, we present the 3D flow of a Carbopol gel in a blocked aperture which resembles more closely the case of fouling in filtration.

In the first phase of this section (i.e.~subsections \ref{sec:Exp2DCavity} \& \ref{num_meth}), we present the experimental and numerical results in regard to the 2D flow over the cavity (i.e.,~Fig.~\ref{fig:setup}b). Both experimental and computational results explore the flow and characterize the position and shape of the yield surfaces. The focus is on the effect of elasticity and plasticity of the fluid in the formation of fouling regions, in particular, the asymmetrical shape of the yield surfaces. In the second phase of this section (i.e.~subsection \ref{sec:3DExp}), we look into experimental results regarding the 3D flow {\color{black}{past}}  %over 
a more complex geometry (i.e.,~Fig.~\ref{fig:setup}c).

Note that in this study, the flow is intertialess and the governing dimensionless groups include Bingham number, which defines the ratio of the yield stress to the characteristic viscous stress,
\begin{equation}
    B = \frac{\tau_y}{K} \left(\frac{H}{U_c}\right)^n,
    \label{bingham}
\end{equation}
and the Weissenberg number, which describes the ratio of the fluid relaxation time ($\lambda$) to the nominal characteristic time of the flow, 
\begin{equation}
    Wi = \frac{\lambda U_c}{L}.
\end{equation}
Please note that $H$ and $L$ are defined in Fig.~\ref{fig:setup} for both geometries. We used $L$ as the characteristic length in the Weissenberg number, as it is a more natural choice since the elastic stresses are dominated by normal stresses which are developed mostly over{\color{black}{/past}} the cavity/aperture~\citep{drost2013hele}.

Under the assumption that Carbopol gels behaves like a Kelvin-Voigt viscoelastic solid below the yielding point, the relaxation time can be estimated by $\lambda=G''/\omega G'$, where $G'$ and $G''$ were measured from oscillatory test in the linear regime (see Table \ref{tab:1}).
%measured during a strain sweep oscillatory test at $\omega=1 \: rads/s$ and $L$ is defined as the characteristic  length or the length of the cavity in this case. %In this work, we asses the plug phenomenology and asymmetry of the YSF in terms of these dimensionless groups. Asymmetry has been qualitatively observed before in flows over cavities, but it was first quantified by Luu et al. as the difference in the physical area of the stagnant regions in the cavity \cite{Philippe2017}. 
%In this paper, however, we follow Chaparian et al. \cite{chaparian2019adaptive} and define asymmetry as the second norm of the velocity difference:

\begin{figure}[h]
    \centering
    \includegraphics[width=10cm]{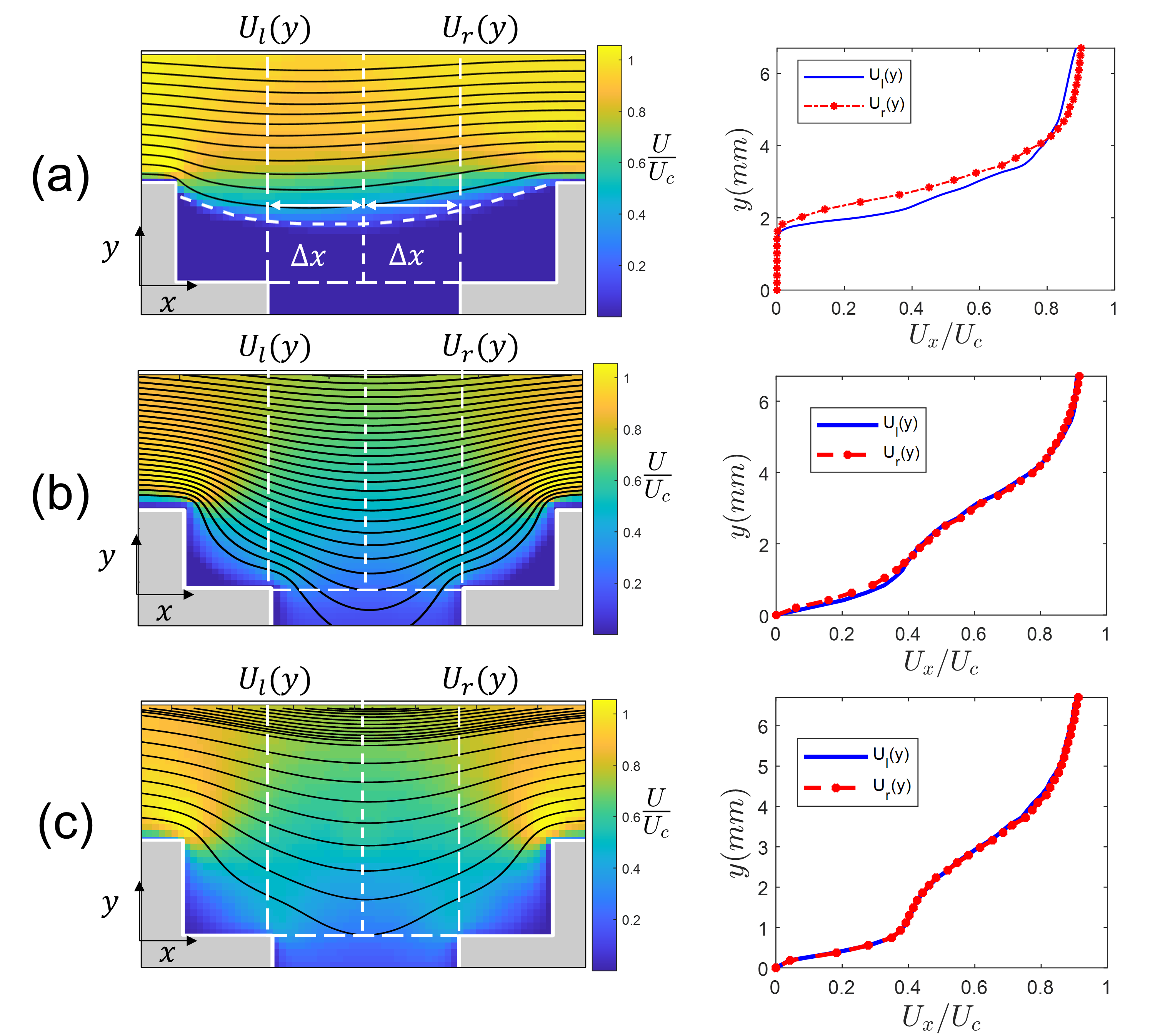}
   \caption{Representative results obtained for the flow of (a) $0.060\%\;(wt/wt)$ Carbopol (b) glycerol solution and (c) PEO solution around an aperture are shown here. The left panels represent the contours of speed, normalised by the far field mean velocity $U_c =Q/WH$, and streamlines on the midplane of the slot around the aperture. In all these cases, the flow rate is fixed at $Q= 1 \:  ml/min$. The extent of asymmetry is defined as the norm of the difference in the $x-$component of the velocity along two symmetric lines separated by the same distance $\Delta x$ from the axis of symmetry of the aperture (see eq.~\ref{asym}). The velocity component in the $x-$direction along those lines, $U_l$ and $U_r$ are plotted in the right panels. Note that the flow of the glycerol solution and PEO solution are symmetrical, while there is a discernible asymmetry in the flow of Catbopol gel which manifests itself in the velocity profiles depicted in (a) and also in the yield surface.    
%   Sample flow fields of (a) $c=0.060\%$ Carbopol (b) glycerol-water and (c) PEO solution are shown on the left. Contours of speed, normalised by the incident speed $U_c =
%Q/WH$, and streamlines (all flowing left to right) along the midplane of the
%slot around the 
%The corresponding profiles of the horizontal component of velocity from the left and right symmetry lines are shown on the right. Note that the glycerol-water and PEO suspensions are symmetrical. The feed flow was set at $Q= 1 \:  ml/min$ in these samples.
   }
    \label{fig:asymdef}
\end{figure}

In this work, we quantify the extent of flow asymmetry by calculating the norm of the velocity difference,
\begin{equation}
    Asymmetry = \int \vert U_r - U_l \vert ~\text{d}y
        \label{asym}
\end{equation}
where $U_r$ and $U_l$ are the dimensionless horizontal velocity profiles along two symmetric lines separated by the same distance $\Delta x$ from the symmetry line of the cavity or aperture. For the cavity $\Delta x=\pm 1 \: mm$ while for the aperture $\Delta x=\pm 3 \: mm$ from the centerline.  
%A unique feature of this definition is that asymmetry based on velocity is independent of the size of unyielded region which is itself dependent on the flow and $B$. 
Fig.~\ref{fig:asymdef} shows a graphical representation of our definition of asymmetry for the flow of a Carbopol gel, glycerol solution and PEO solution at a fixed flow rate. The figure includes the flow fields (left panel) and their corresponding horizontal velocity profiles (right panel) along the chosen lines. Note that the flow  profiles measured along these two lines are identical in the case of the glycerol and PEO solutions (see Fig.~\ref{fig:asymdef}b-c), whereas there is a clear difference between them for the Carbopol gel (see Fig \ref{fig:asymdef}a). We discuss further this flow asymmetry in what follows.

\subsection{Plug Phenomenology and Asymmetry in Cavity Flows}\label{sec:Exp2DCavity}
In this section, the flow dynamics and the static plug morphology are characterized for the flow of Carbopol over a cavity by means of particle image velocimetry discussed in section \ref{Sec:process_meth}.

\begin{figure}
    \centering
    \includegraphics[width=7 cm]{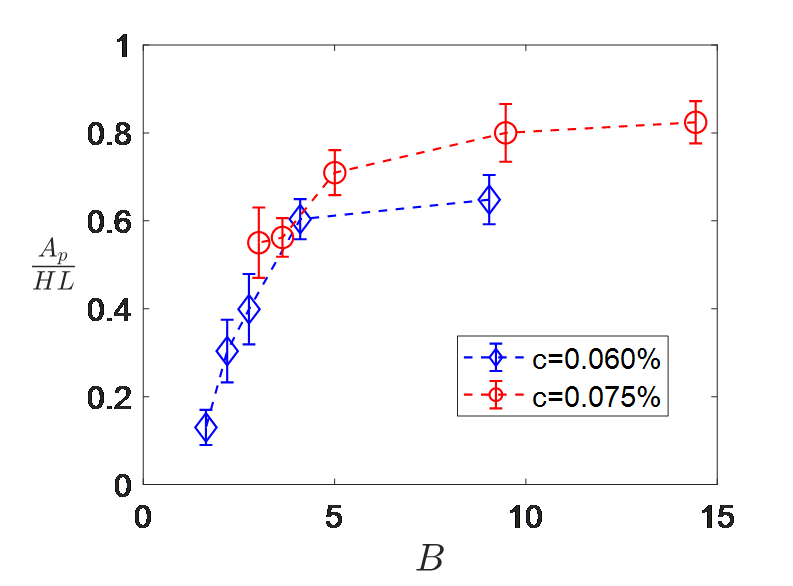}
    \caption{
    Variation of the dimensional area of the plugged region $A_p$ developing inside the 2D cavity normalized by the area of the cavity (i.e. $H L$) plotted as function of $B$. The red and blue symbols represent the data obtained for $0.060\%\;(wt/wt)$ and $0.075\%\;(wt/wt)$ Carbopol gels, respectively. {\color{black} The data in the figure was obtained from averaging  repeated measurements of the area under the yield surfaces using two methods, namely the light intensity thresholding algorithm for the streakline images and the noise-based thresholding for the velocimetry measurements. The error bars represent the standard deviation.}
    }
    \label{fig:plugarea2D}
\end{figure}

The static yield surfaces are located always inside the gap regardless of the average velocity and Carbopol concentration (see Fig.~\ref{fig:plug2D}). As expected, at an asymptotically small flow rates the yield surfaces approach the channel bottom wall (i.e.~the top of the cavity). However, we observe an erosion of the plugged region by either increasing the velocity or decreasing the gel concentration (i.e.~the yield stress). This suggests that the position of yield surfaces and the size of the plugged region developing in the cavity can be formulated as a function of the Bingham number. Fig.~\ref{fig:plugarea2D} presents the dimensionless size of the plug, i.e. the ratio of the area of the plugged region ($A_p$) to the cavity area ($HL$), versus the Bingham number for the flow with different average velocities and Carbopol with two different concentrations. {\color{black} The data in the figure is obtained from averaging repeated measurements of the area under the yield surface using the two methods outlined in section \ref{Sec:process_meth}}. As demonstrated from this figure, the plug area grows by increasing the Bingham number.
%For a given Carbopol suspension, the position of the interface changes with the flow rate. As the feed flow $U_c$ is increased, the interface is displaced closer to the cavity wall and the area of the fouling layer or plug is reduced. The plug displacement or erosion is greater for the lower concentration Carbopol which naturally has a lower yield stress and in our experiments this difference becomes much greater as the flow rate increases. These observations are quantitatively demonstrated in terms of plug area in Fig \ref{fig:plugarea2d}a. Consistently, the plug area is lower for the $c=0.060 \%$ Carbopol compared to the $c=0.075 \%$ Carbopol at the same characteristic velocity. Plug area as function of Bingham number is shown in Fig \ref{fig:plugarea2d}b. Interestingly, when plotted against $B$, we note that the plug area data for the two Carbopol gels roughly collapse on to a single curve as $B$ decreases. This implies that the are of plug is a function of $B$, i.e. yield stress and $U_c$. Similar behavior was observed when Daneshi et al. plotted plug length as function of $B$ for viscoplastic flows around obstacles \cite{Daneshi2020}. 
%Theoretically speaking, this occurs because the effects of shear-thinning index $n$ in Eq \ref{bingham} becomes negligible and the plug size becomes only a function of the $B$ number as $B$ becomes small. Conversely, as $B$ becomes larger, the yield stress dominates over the viscous stress, thus the difference between the two Carbopol gels becomes more apparent. 
{\color{black}Interestingly, when plotted against $B$, we note that the plug area data for the two Carbopol gels  show a degree of overlap, in particular for $B<10$.  This implies that the area of plug is a function of $B$ and does not noticeably change by the power-law index of the gel.}
%Interestingly, when plotted against $B$, we note that the plug area data for the two Carbopol gels roughly collapse on a single curve. This implies that the area of plug is only a function of $B$ and does not noticeably change by the power-law index of the gel.
Similar behavior was observed by Daneshi et al. \cite{Daneshi2020} for Carbopol flow around obstacles considering the plug length as a function of $B$.

Now, we turn our attention to the morphology of the plugs. In particular, we note asymmetry in the static plug regions of the flow of Carbopol gels over the cavity, where the yield surfaces drop deeper on the left half of the cavity. Remarkably, the flows in the experiment are virtually inertialess and, as such, one would expect a symmetric flow for viscoplastic materials. However, the flow asymmetry is clearly visible across the entire range of $B$ and $Wi$ numbers presented here for Carbopol samples whereas the flow was entirely symmetric for the PEO and glycerol solutions (see Fig.~\ref{fig:asymdef}b-c). 

\begin{figure}[!h]
    \centering
    \includegraphics[width=10 cm]{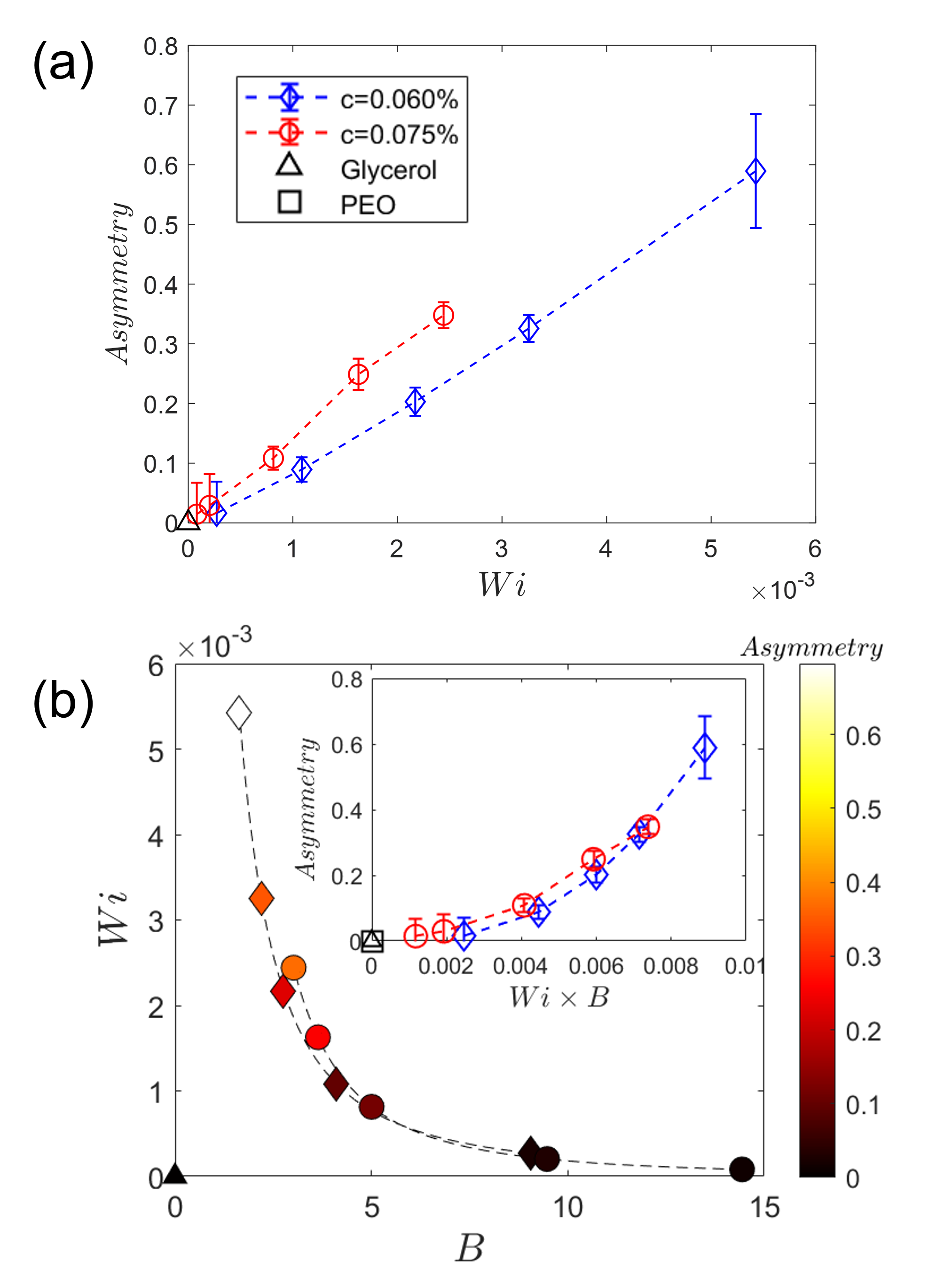}
    \caption{The quantified asymmetry of the 2D flow over the cavity (Fig.~\ref{fig:setup}b). The results were reported for the flows of a glycerol solution, a PEO solution, and $0.060\%\;(wt/wt)$ \& $0.075\%\;(wt/wt)$ Carbopol gels. Panel (a) shows the extent of asymmetry as a function of $Wi$. Panel (b) recasts the data of (a) in $Wi-B$ panel where the extent of asymmetry is presented by the color of the symbols. The extent of asymmetry versus $Wi \times B $ is plotted in the inset of panel (b), where the data obtained for both Carbopol gels collapse into a master curve. Note that the extent of asymmetry calculated for PEO solution is less than $\mathcal{O}(10^{-3})$, while the $Wi$ estimated for this material is around $0.075$ which is an order of magnitude larger than those for Carbopol gels and lie beyond the range of axes in panel (a) and (b), respectively. {\color{black}The error bars include the uncertainties from both the visualization resolution and the asymmetry variation from moving the symmetry lines (see Fig.~\ref{fig:asymdef}) by $\pm 10\% \Delta x$.}}
    \label{fig:ASYM2D}
\end{figure}

The extent of asymmetry (calculated by Eq.~\ref{asym}) for Carbopol as a function $Wi$ is shown in Fig.~\ref{fig:ASYM2D}a. The error bars displayed in the %figure
{\color{black} asymmetry measurements} include the uncertainties from both the visualization resolution and the asymmetry variation from moving the symmetry lines (see Fig.~\ref{fig:asymdef}) by $\pm 10\% \Delta x$. Evidently from this figure,  the extent of asymmetry becomes more pronounced by increasing $Wi$, similar to what was reported for the viscoelastic polymers in the literature \cite{kim2000instabilities,yamamoto2003}. However, the surprising feature of these results is that the asymmetry is significant even at very small $Wi$ numbers where for a viscoelastic fluid we used, PEO solution, the flow is symmetrical.

We also note that at a fixed $Wi$, the flow is more asymmetric for the $c =0.075 \%\;(wt/wt)$ Carbopol than that of the $0.060\%\;(wt/wt)$ Carbopol. These observations imply that, in addition to the elasticity (i.e.~Weissenberg number), the yield stress of the material plays a role. To better understand the role of these dimensionless numbers, we plot the asymmetry on $Wi-B$ plane with the extent of asymmetry is shown by color of each symbol (see Fig.~\ref{fig:ASYM2D}b). The extent of asymmetry decreases as we go to smaller Weissenberg numbers and larger Bingham numbers (bottom right of the $Wi-B$ plane in panel (b)) for both Carbopol gels. Encouraged by the suggestion of Chaparian \& Tammisola \cite{chaparian2019adaptive}, we plot the extent of asymmetry versus $Wi \times B$ in the inset of Fig.~\ref{fig:ASYM2D}b. As it can be seen, the data fit satisfactorily on a master curve with this new scaling.This insinuates that $Wi \times B$ which accounts for both elastic and plastic behaviour of the material can be used to explain the asymmetry. This holds valid for the solutions without yield stress, i.e. glycerol solution and PEO, as well. Note that although PEO solution is known as a viscoelastic fluid, the lack of asymmetry in the flow of PEO is attributed to the small Weissenberg numbers in the range of flow rates in our experiments. These results suggest that the asymmetry is triggered by the yield stress of the fluid at low $Wi$ numbers. 

% {\color{black}(Miguel: should we say the what is the Wi number in this case? I calculated it to be Wi =0.025 for PEO. I found in papers of viscoelastic flows over cavities that the symmetry is broken at roughly Wi = 0.3.)}
%
%It was shown that the elastic strain is maximum along
%the yield surfaces and changes the shape of the unyielded
%region within the cavity. This observation is in accordance
%with experimental observations reported in the literature. It
%is also shown that the plug flow in the upstream and downstream
%channels is suppressed in the vicinity of the cavity as
%a result of the elasticity effects.
%The effect of viscoplasticity was
%
%Finally, it was found that the elastic extra-stress field is rather
%asymmetric, with a high stress zone near the expansion plane,
%which causes the enlargement of the yielded region in the vicinity
%of the cavity inlet.
%
%displaying a high stress zone near
%the expansion plane that has no counterpart near the contraction
%plane. The existence of this high stress zone is certainly responsible
%for the enlargement of the yielded region in the vicinity of the cavity
%inlet.

%\begin{figure}[]
%    \centering
%    \includegraphics[width=10 cm]{FIGURESAPRIL2022/asymmetry_cavity.png}
%    \caption{Asymmetry of YSF flows over a quasi 2D cavity. Asymmetry as function of (a) Bingham number $B$ and (b) Weissenberg numbers $Wi$. $Wi$ as function of $B$ is plotted in (c), where the contour quantifies the extent of asymmetry. Asymmetry as function of $Wi \times B $ plotted in (d).}
%    \label{fig:ASYM2D}
%\end{figure}

\subsection{Numerical simulation of the 2D flow over cavity using elastoviscoplastic rheological model}\label{num_meth}

In this study, we simulate the Stokes flow in the 2D cavity (Fig.~\ref{fig:setup}b) using the same numerical scheme proposed by Chaparian \& Tammisola \cite{chaparian2019adaptive} for an elastoviscoplastic rheological model introduced by Saramito \cite{SARAMITO2007}:
\begin{equation}\label{eq:constitutive}
\underbrace{\lambda ~\overset{\triangledown}{\boldsymbol{\uptau}} }_{\dot{\boldsymbol{\upgamma}}_e} + \underbrace{ \left( 1-\frac{\tau_y}{\Vert \boldsymbol{\uptau} \Vert_{\text{v}}} \right)_+ \boldsymbol{\uptau} }_{\dot{\boldsymbol{\upgamma}}_p}=  \mu \dot{\boldsymbol{\upgamma}},
\end{equation}
{\color{black}
or in the dimensionless form,
\begin{equation}\label{eq:constitutive_dimensional}
Wi ~\delta ~\overset{\triangledown}{\boldsymbol{\uptau}^*} + \left( 1-\frac{B}{\Vert \boldsymbol{\uptau}^* \Vert_{\text{v}}} \right)_+ \boldsymbol{\uptau}^* =  \dot{\boldsymbol{\upgamma}}^*,
\end{equation}
where $\delta=H/L$, $\Vert \cdot \Vert_{\text{v}}$ is the von Mises criterion (i.e.~based on the second invariant of the stress tensor), $\left( \cdot \right)_+$ is the positive part of the argument (i.e. is equal to zero if the argument is negative and otherwise is equal to the argument), and $( \overset{\triangledown}{\cdot} )$ is the upper-convected derivative. Please note that in this rheological model $n=1$. Indeed, after yielding, the material behaves as a Bingham fluid with an extra elastic memory.}

This computational method is based on splitting the entire problem into viscoelastic and viscoplastic parts. The viscoplastic subproblem (i.e.~$\dot{\boldsymbol{\upgamma}}_p$ part in the constitutive equation) is handled by the augmented Lagrangian method \cite{roquet2003adaptive}. Then, the two subproblems are superimposed (i.e.~$\dot{\boldsymbol{\upgamma}}_e+\dot{\boldsymbol{\upgamma}}_p$) until the equation (\ref{eq:constitutive}) is satisfied up to the numerical convergence.

We use an open source finite element environment FreeFEM++ \cite{MR3043640} for discretization and meshing which has been %widely 
validated in our previous studies \cite{chaparian2017yield,iglesias2020computing,chaparian2020sliding}. Anisotropic adaptive mesh is combined with this method to get smoother yield surfaces and ensure high resolution of the flow features \cite{chaparian2019adaptive,roquet2003adaptive}. We shall mention that to avoid numerical instabilities, although the simulated geometry is exactly the same as Fig.~\ref{fig:setup}b, we slightly round the sharp corners and carefully monitor that the effect is locally restricted. Note that the level of asymmetry is measured according to the same metric introduced at the beginning of Section \ref{sec:results}.

\begin{figure}[h]
    \centering
    \includegraphics[width=\textwidth]{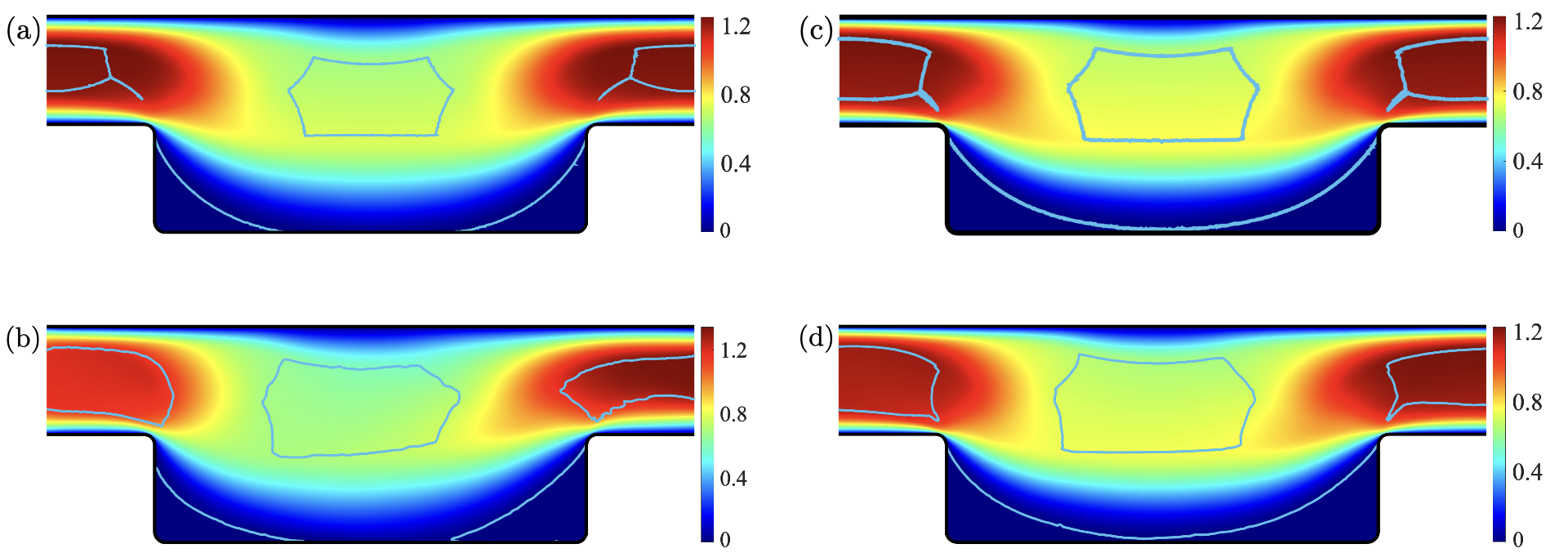}
    \caption{Velocity contour for the case : (a) $B=10, Wi=0$; (b) $B=10, Wi=0.0125$; (c) $B=20, Wi=0$; (d) $B=20, Wi=0.0025$. The yield surfaces are shown in cyan.}
    \label{fig:NumericalVelocityContour}
\end{figure}

Four sample computations are shown in Fig.~\ref{fig:NumericalVelocityContour}; two viscoplastic simulations (i.e.~$Wi=0$ in panels (a) \& (c)) and two elastoviscoplastic cases. As can be observed, by increasing the Bingham number, the unyielded plugs grow which is intuitive. {\color{black}{The important point here is that there is a discernible asymmetry in the flow field of the elastoviscoplastic cases although the geometry is completely symmetrical. This asymmetric flow is attributed to the elasticity of the fluid.}}
%The important point here is that the asymmetry in the flow field of elastoviscoplastic counterparts although the geometry is completely symmetric which is attributed to the elasticity of the fluid. 
Please note that the Weissenberg numbers are small in both cases.

\begin{figure}[!h]
    \centering
    \includegraphics[width=\textwidth]{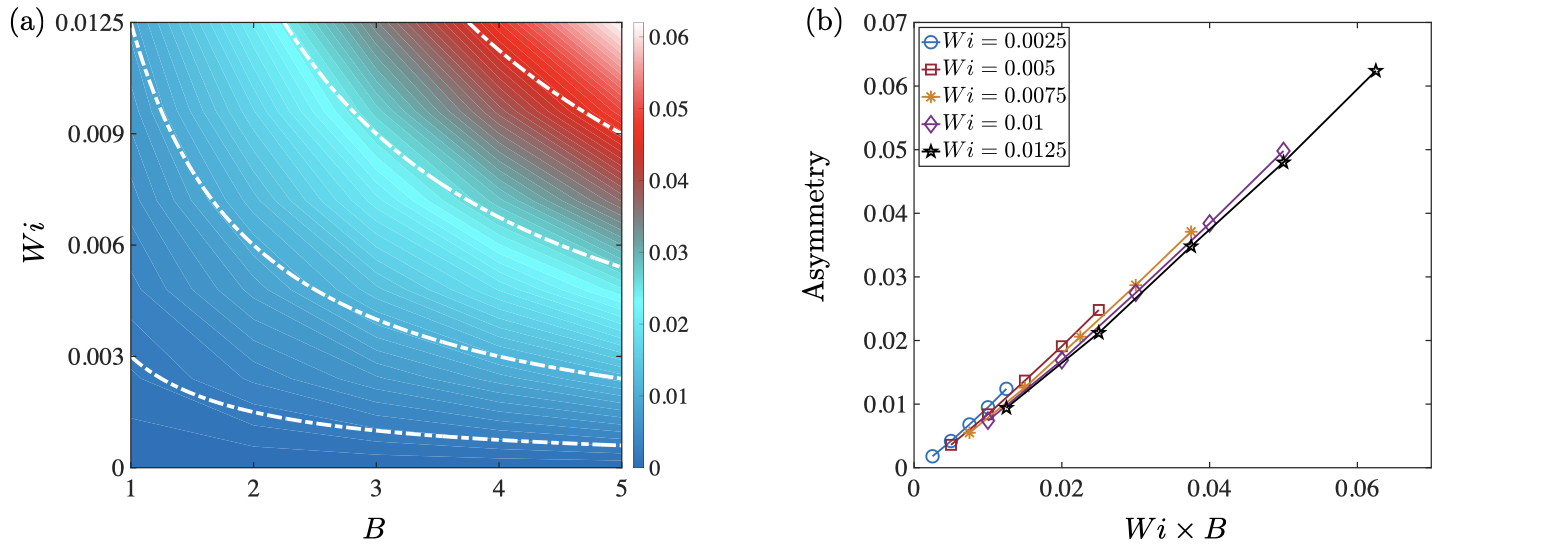}
    \caption{Results of the numerical simulation where (a) shows the asymmetry contour in Weissenberg-Bingham plane and (b) shows asymmetry versus $Wi \times B$.}
    \label{fig:AsymmetryContour}
\end{figure}

 We visualized the asymmetry in two different ways in Fig.~\ref{fig:AsymmetryContour} by spanning the Weissenberg and the Bingham numbers up to 0.0125 and 5, respectively, and by plotting as function of $Wi \times B$. Hence, the contours in panel (a) and the $y-$axis of panel (b) represent the level of asymmetry. The white curves sketched on top of the contours in panel (a) represents the $Wi \times B = const.$ lines. Some general points can be drawn from this figure. Perhaps the most important one is that at a fixed Weissenberg number, if we increase the Bingham number, the flow will be more and more asymmetric. In other words, plasticity triggers the elastic effects as well. This is in the same direction as the conclusions made by Chaparian and co-workers in a series of studies in various physical problems from elastoviscoplastic fluid flows in porous media and complex geometries \cite{chaparian2019porous,chaparian2020particle} to particle migration when the carrier fluid is elastoviscoplastic \cite{chaparian2020particle}. This is also consistent with our experimental results (see Fig.~\ref{fig:ASYM2D}b). Another interesting point is that the level of asymmetry is almost the same when $Wi \times B$ is constant: please see all the curves in Fig.~\ref{fig:AsymmetryContour}b almost coincide on top of each other. Please note that the level of asymmetry grows almost linearly with $Wi \times B$. Similar trend was observed in the experimental results discussed in the previous subsection. 

%---------------------------------------------------------------------------%

\subsection{Plug Phenomenology and Asymmetry in a Blocked Aperture}\label{sec:3DExp}

%In this section, we visualize the flow of Carbopol gels in a blocked aperture, i.e. the flow is constrained not by a wall but by the yield stress of the fluid. This flow is similar in geometry to the flow over a cavity. However, the yielded area can drop below the step into the aperture. This flow is analogous to the simplest case in cross-flow filtration where the aperture is blocked. The raw images of flow and the velocity contour for this geometry are displayed in Fig \ref{fig:plug3d}a for the $c=0.060 \%$ gel and in Fig \ref{fig:plug3d}b for the $c=0.075 \%$ gel. By inspection, we see that the size of the plug zones is reduced with increasing flow $U_c$ (see Fig \ref{fig:plugarea3d}a). Intuitively, it is also evident that the higher concentration Carbopol has bigger plug zones at the same flow $U_c$. Fig \ref{fig:plugarea3d}b shows the plug area as function of Bingham number $B$. When plotting plug area as function of $B$ we obtain a similar shape curve compared to the flow over cavity. Notably, the collapse of the data also improves when plug area is plotted against $B$. This implies that the extent of plug erosion is a function of yield stress. 

\begin{figure}[!h]
    \centering
    \includegraphics[width=13 cm]{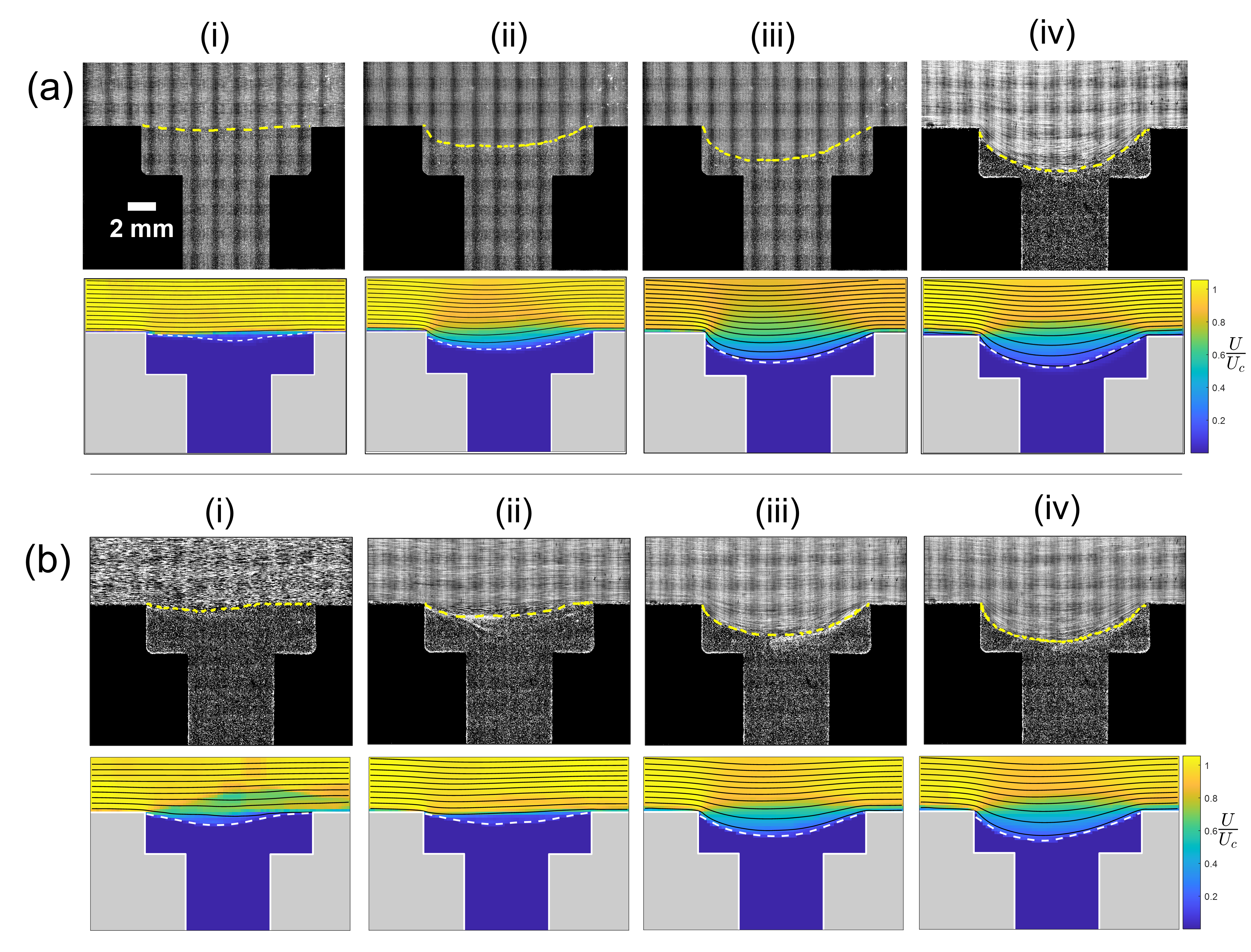}
    \caption{Representative results regarding the flows of (a) $0.060\%\;(wt/wt)$ and (b) $0.075\%\;(wt/wt)$ Carbopol gels {\color{black}{past}} %over
an aperture. The top series of each panel shows the streaklines flow and the bottom series of each panel shows the corresponding normalized speed contour ($U/U_c$). The yellow and white dashed lines highlights the yield surfaces. %The series (i-iv) shows the change in flow corresponding to $U_c=0.01,0.10,0.20,0.28 \ mm/s$. 
     From left to right (from $(i)$ to $(iv)$) the upstream velocity increases from $0.01$ to $0.28\;mm/s$. Hence, they correspond to (a): $0.060\%\;(wt/wt)$ Carbopol gel, (i) $B = 7$, (ii) $B = 1.8$ and (iii) $B = 1.3$ and (iv) $B =1.0$; and to (b): $0.075\%\;(wt/wt)$ Carbopol gel (i) $B = 7.7$, (ii) $B = 2.6$ and (iii) $B = 1.9$ and (iv) $B =1.6$.}
    \label{fig:plug3d}
\end{figure}

Here, we present the results regarding the flow of Carbopol gels {\color{black}{past}} %over 
a blocked aperture. 
%Similar to the flow over cavity, an unyielded region forms in the aperture where its size changes by the mean velocity in the channel and Carbopol concentration. 
{\color{black}{Similar to what was observed for the flow over cavity, an unyielded region forms in the aperture with its size changing with the mean velocity in the channel and the Carbopol concentration}}.
%{\color{black}{Similar to the flow over a cavity, an unyielded region forms in the aperture. The size of this region changes depending on the mean velocity in the channel and the Carbopol concentration}}.
However, the flow in this geometry is different from the 2D flow over the cavity: here, the flow is three-dimensional and the shear stresses along the depth of the channel are significant. The raw images of the flow and the velocity contours in this geometry are displayed in Fig.~\ref{fig:plug3d}a for the $0.060 \%\;(wt/wt)$ Carbopol gel and in Fig.~\ref{fig:plug3d}b for the $c=0.075 \%\;(wt/wt)$ gel. The static yield surfaces develop in the aperture which always lie below the span of the channel wall.
{\color{black}Additionally, in the range of our experiments, the yield surface lies above the slot below the main cavity. In other words, the slot is {\it cloaked} (i.e.~is hidden) inside an unyielded envelope. This is reminiscent of the ``cloaking effect" explained in \citep{Daneshi2020,chaparian2017cloaking}, implying that, in our experiments, the unyielded plug inside the slot masks the actual shape of the apparatus. Hence, there is no difference between the flow in this geometry and the geometry which only has the main cavity.}

\begin{figure}[!h]
    \centering
    \includegraphics[width=8 cm]{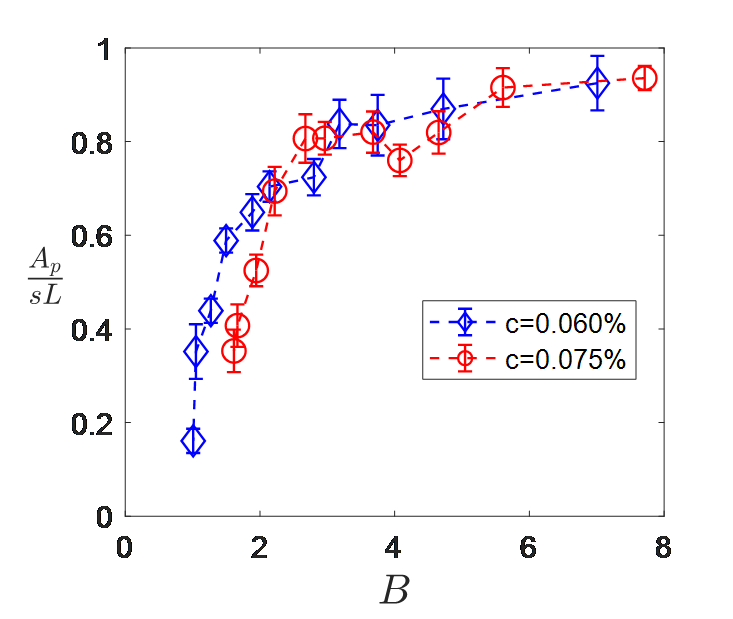}
    \caption{The dimensional area of plugged region $A_p$ developing inside the aperture normalized by the area of the step of the aperture (i.e.~$s L$) versus $B$. The red and blue symbols represent the data obtained for $0.060\%\;(wt/wt)$ and $0.075\%\;(wt/wt)$ Carbopol gels, respectively. {\color{black} The data in the figure was obtained from averaging  repeated measurements of the area under the yield surfaces using two methods, namely the light intensity thresholding algorithm for the streakline images and the noise-based thresholding for the velocimetry measurements. The error bars represent the standard deviation.}
%Plug area normalized by the step area of the aperture as function of Bingham number $B$. The error bar arises from the size of interrogation window in the velocity measurements and the size of pixels.
%    Plug area normalized by the step area of the aperture as function of Bingham number $B$. The error bar arises from the size of interrogation window in the velocity measurements and the size of pixels. 
    }
    \label{fig:plugarea3d}
\end{figure}

As expected, at higher flow rates, the plugged regions are eroded more and the unyielded regions shrink. Intuitively, it is also evident that the higher concentration Carbopol has larger plug zones at the same flow rate. Fig.~\ref{fig:plugarea3d} shows the plug area $A_p$ normalized by the area of the aperture step ($sL$) as a function of the Bingham number. Again here, when plotting plug area as a function of $B$, we obtain %a collapse 
{\color{black} a good degree of overlap}
of the data for the two different gels. This is in line with the results obtained for the 2D flow over cavity and implies that the plug size can be explained as a function of Bingham number and does not change noticeably by the degree of shear thinning of the material.  

%\begin{figure}[]
%    \centering
%    \includegraphics[width=13 cm]{FIGURESAPRIL2022/plugerosion0060_2.png}
%    \caption{Streaklines (bottom) and velocity contour (top) for the flow of $c= 0.060 \%$ Carbopol over an aperture. The flows of $U_c=0.01-0.28 \ mm/s$ corresponds to $B=7.0-1.1$, respectively. The yellow dashed line highlights the yielding surfaces. Color bar represent normalized velocity values.}
%    \label{fig:plug3d1}
%\end{figure}
%
%
%\begin{figure}[]
%    \centering
%    \includegraphics[width=13 cm]{FIGURESAPRIL2022/plugerosion0075_2.png}
%    \caption{Streaklines (bottom) and velocity contour (top) for the flow of $c= 0.075 \%$ Carbopol over an aperture. The flows of $U_c=0.01-0.28 \ mm/s$ corresponds to $B=7.7-1.6$, respectively. The yellow dashed line highlights the yielding surfaces. Color bar represent normalized velocity values.}
%    \label{fig:plug3d}
%\end{figure}

%\begin{figure}[]
%    \centering
%    \includegraphics[width=13 cm]{FIGURESAPRIL2022/aperture_area.png}
%    \caption{Plug area normalized by the step area as function of (a) upstream characteristic velocity $U_c$ and (b) Bingham number $B$.}
%    \label{fig:plugarea3d}
%\end{figure}

\begin{figure}[!h]
    \centering
    \includegraphics[width=10 cm]{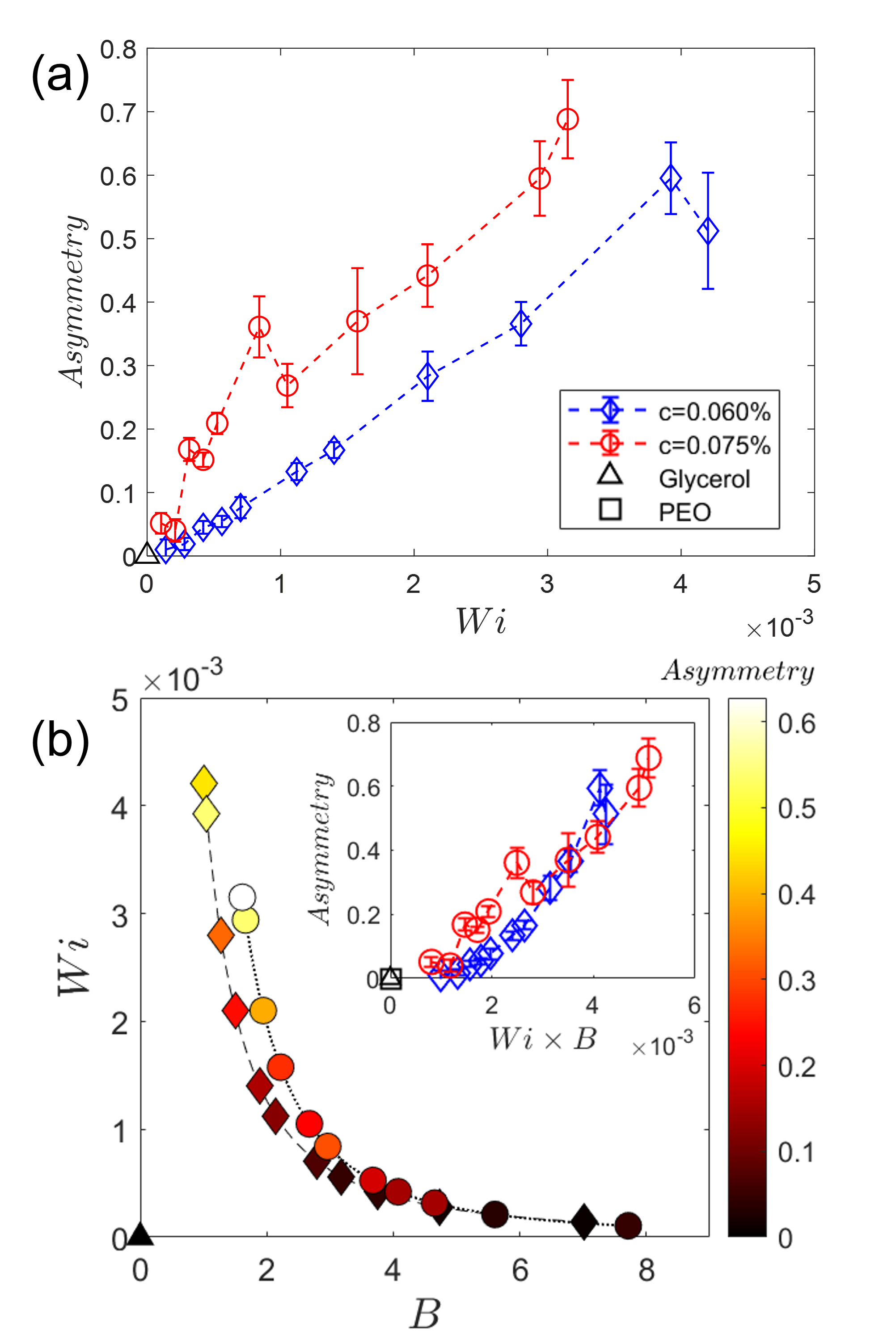}
   \caption{The extent of asymmetry associated with the 3D flow inside the aperture (Fig.~\ref{fig:setup}c). The results are reported for the flows of a glycerol solution, a PEO solution, and $0.060\%\;(wt/wt)$ \& $0.075\%\;(wt/wt)$ Carbopol gels. Panel (a) shows the extent of asymmetry as a function of $Wi$. Panel (b) recasts the data of (a) in $Wi-B$ panel where the extent of asymmetry is presented by the color of the symbols. The extent of asymmetry versus $Wi \times B $ is plotted in the inset of panel (b), where the data obtained for both Carbopol gels collapse into a master curve. The extent of asymmetry calculated for PEO solution is around $\mathcal{O}(10^{-3})$, hence, similar to Fig.~\ref{fig:ASYM2D}, here the data obtained for PEO solution is absent in panels (a) and (b).{\color{black}The error bars include the uncertainties from both the visualization resolution and the asymmetry variation from moving the symmetry lines (see Fig.~\ref{fig:asymdef}) by $\pm 10\% \Delta x$.}     
   }
    \label{fig:asym3d}
\end{figure}

Similarly to the 2D flow over the cavity, we observe that flow asymmetries clearly increase by $Wi$ (see Fig.~\ref{fig:asym3d}a). This observation further highlights the robustness of the asymmetry in the flows of Carbopol gels over non-smooth geometries regardless of the flow type. Fig.~\ref{fig:asym3d}a and Fig.~\ref{fig:asym3d}b confirms this observation quantitatively. The higher concentration Carbopol has more asymmetric flow and the difference in flow asymmetries between the two Carbopol gels is greater in the aperture than in the flow over cavity. %The difference could be likely due to the different geometries; the former being categorically a 3D channel while the latter is more closely approaching the Hele-Shaw limit.
In tandem with the 2D flow over the cavity, in Fig \ref{fig:asym3d}b, we see that asymmetry increases at a fixed $Wi$ with the Bingham number. %Similar results have been shown by numerical simulations of flows in wavy channels \cite{chaparian2019adaptive} and in our numerical and experimental results for the 2D flow over a cavity.
The extent of asymmetry as a function of $Wi \times B$ is shown in the inset of Fig \ref{fig:asym3d}b for the aperture geometry. When plotted against  $Wi \times B$, the data obtained for both Carbopol gels lie on each other again --- the same as what we have observed for for the 2D flow over a cavity.

%\begin{figure}[]
%    \centering
%    \includegraphics[width=14 cm]{FIGURESAPRIL2022/asymcomp_aperture.png}
%   \caption{Asymmetry of YSF flows over an aperture. Asymmetry as function of (a) Bingham number $B$ and (b) Weissenberg numbers $Wi$. $Wi$ as function of $B$ is plotted in (c), where the contour shows asymmetry values. Asymmetry as function of $Wi \times B $ plotted in (d).}
%    \label{fig:asym3d}
%\end{figure}

\section{Concluding Remarks}\label{sec:conc}

In this study, we revisited a classic fluid mechanics problem, flow in a conduit with an abrupt expansion-contraction. Due to the yield stress of the material, an unyielded region fills inside the cavity. The position and shape of the yield surface and the flow characteristics in its vicinity are scrutinized in this work, from both experimental and numerical perspectives.

In the first phase of this study, we examined the two-dimensional flow over a cavity. A high resolution micro-PIV measurements were conducted to determine the flow field developing around the cavity and the position of the yield surface. We also complemented these experiments with computations (i.e.~section \ref{num_meth}) developed based on an elastoviscoplastic constitutive law describing the rheology of the fluid. The experiments show the growth of the dead zones developing in the cavity with the Bingham number which is intuitive. A through analysis of the results reveals the asymmetry of the yield surfaces and the flow field developing around the cavity. The surprising feature of our findings is that the asymmetry is absent in the {\color{black} \it relatively} low Weissenberg number flow of a viscoelastic fluid, PEO solution, while there is a markedly noticeable level of asymmetry for the flow of elastoviscoplastic fluid, Carbopol gel, {\color{black} not only in the similar range of the Weissenberg number, but also at smaller values.} These observations are robust in the computations as well, confirming the existence of asymmetry in the low Weissenberg number flow of elastoviscoplastic fluids. This implies that a combination of elastic and plastic effects magnify the level of asymmetry.
%{\color{blue}COMMENT: MAYBE BETTER TO REMOVE?? Note that the flow in our experiments is very slow and the $Wi$ number is very small. Hence, it is expected for the flow of viscoelastic fluids to be symmetrical. However, asymmetrical flow is developed as the yield stress comes to play.}

The experiments were repeated in a more complex geometry, where there is a three-dimensional flow {\color{black}{past}} %over
 an aperture in a thin slot. In contrast with the previous geometry, in this case, the stresses are not restricted into 2D. Indeed, the confinement in the $z-$direction (see Fig.~\ref{fig:setup}b) dictates significant shear stress to the fluid and cannot be neglected compared to the normal and shear stresses developed in the $x-y$ plane across the aperture. Despite these differences, similar trends in the flow asymmetries have been observed in this case which implies that the asymmetry scalings presented is universal and are not dependent on geometry, Carbopol concentration, or flow rates. %is robust, as demonstrated by varying the geometry (cavity/aperture), concentration of Carbopol gels, and flow rates. 

Asymmetry has been known as a characteristic of viscoelastic fluid flows in these types of geometries, where $Wi$ number is sufficiently high. In particular, the asymmetry observed for the flow of viscoelastic fluids over a cavity is reminiscent of the ``die swelling" effect \cite{kim2000instabilities} and can be attributed to the viscoelastic relaxation of the stresses developing as the fluid passes over the cavity. The elastic strains develop as the fluid elements pass through the expansion. This leads to an asymmetric stress field where a high stress zone forms near the expansion in contrast to the flow down-stream in the vicinity of the contraction. {\color{black} In viscoelastic fluids, asymmetry is a function of the Weissenberg number, and goes to zero as Wi approaches zero.}

The scenario is more complex for the flow of yield-stress fluids where asymmetry exists even at very low $Wi$ numbers {\color{black} where a pure viscoelastic fluid's asymmetry is not measurable in the experiments and is
an order of magnitude less in the computations; see Fig.~\ref{fig:AsymmetryContour}.} To characterize and delve into the anomalous asymmetry observed for low $Wi$-number flow of yield-stress fluids in these geometries, we quantified the asymmetry by measuring the average difference in the velocity of the fluid passing the geometry symmetry line. As preliminary introduced by Chaparian \& Tammisola \cite{chaparian2019adaptive}, when the level of asymmetry is plotted against $Wi \times B$, a collapse of data is achieved. This feature is obtained not only for the experimental and computational data regarding the two-dimensional flow over the cavity but also for the experimental measurements concerning the 3D flow %over 
{\color{black}{past}}
an aperture. It suggests that $Wi \times B$ could be used to quantify the elastic behaviour of an elastoviscoplastic fluid rather than the $Wi$ number. In other words, an elastoviscoplastic fluid manifests more elastic characteristic than its viscoelastic counterpart. Indeed, the plasticity of the fluid triggers the elastic behaviour. This fact has been reported previously in the context of flow over a single or arrays of obstacles \cite{cheddadi,Daneshi2020,chaparian2019porous}, flow through a wavy channel \cite{chaparian2019adaptive}, and particle migration in elastoviscoplastic fluids \cite{chaparian2020particle}.

This aspect can be more understood by reviewing the nature of the $Wi$ and $B$ numbers. The viscoelastic relaxation time $\lambda$ is associated with the necessary time for the stress of the fluid to decay when the motion is brought to a halt. In this case the viscosity is responsible for the work dissipation due to stress relaxation. Hence, for a viscoelastic fluid, the $Wi$ number is the ratio of the relaxation time to the viscous hydrodynamic time scale: $Wi = \lambda \tau_c / \mu = \lambda U_c / \ell$ where $\tau_c$ is the characteristic viscous stress. However, in the case of an elastoviscoplastic material, not only does the viscosity dissipate the stored elastic energy, but also plasticity of the material would contribute in the stress relaxation. Therefore, the ``effective" $Wi$ number for an elastoviscoplastic fluid increases with the plasticity or indeed the Bingham number. In other words, $Wi \times B = \lambda ~\tau_y / \mu$ which is the ratio of the relaxation time over the ``viscoplastic" time scale seems more reasonable to characterize how elastic the material is.

A lot still remains to be understood about the complex behaviours of elastoviscoplastic materials and the augmentation of elasticity and plasticity in yield-stress fluids. Here, we shed some light on of the important class of yield-stress fluid flows, i.e.~internal flows over non-smooth geometries which has a lot of applications in various industries, e.g.~filtration. The present study provides some base fundamental scalings from which to further the knowledge about elastoviscoplastic materials at more details in complex hydrodynamic flows and/or the rheological behaviour of practical yield-stress fluids near the yielding point.

%Here, we focus on the generalization of this problem to the situation in which a practical yield stress fluid flows in a thin slot containing a cavity or an aperture.

%Viscoelastic fluid flows over a cavity exhibit the same characteristic which is related to ``die swelling" effect.}


\section*{Acknowledgment}
The financial support through the NSERC Collaborative Research program, in conjunction with Advanced Fibre Technologies, is gratefully acknowledged.


%\section*{References}


\bibliographystyle{elsarticle-num} 
  %\bibliography{library}
\bibliography{library_II}

\begin{thebibliography}{10}
\expandafter\ifx\csname url\endcsname\relax
  \def\url#1{\texttt{#1}}\fi
\expandafter\ifx\csname urlprefix\endcsname\relax\def\urlprefix{URL }\fi
\expandafter\ifx\csname href\endcsname\relax
  \def\href#1#2{#2} \def\path#1{#1}\fi

\bibitem{BHAVE2014149}
C.~C. Todaro, H.~C. Vogel, Fermentation and biochemical engineering handbook,
  William Andrew, 2014.

\bibitem{Ovarlez2013}
G.~Ovarlez, S.~Cohen-Addad, K.~Krishan, J.~Goyon, P.~Coussot, On the existence
  of a simple yield stress fluid behavior, J. Non-Newtonian Fluid Mech. 193
  (2013) 68--79.

\bibitem{Derakhshandeh2012}
B.~Derakhshandeh, D.~Vlassopoulos, S.~G. Hatzikiriakos, Thixotropy, yielding
  and ultrasonic {D}oppler velocimetry in pulp fibre suspensions, Rheol. Acta
  51~(3) (2012) 201--214.

\bibitem{Emady2013}
H.~Emady, M.~Caggioni, P.~Spicer, Colloidal microstructure effects on particle
  sedimentation in yield stress fluids, J. Rheol. 57~(6) (2013) 1761--1772.

\bibitem{ROUSTAEI2013109}
A.~Roustaei, I.~A. Frigaard, The occurrence of fouling layers in the flow of a
  yield stress fluid along a wavy-walled channel, J. Non-Newtonian Fluid Mech.
  198 (2013) 109--124.

\bibitem{Daneshi2020}
M.~Daneshi, J.~MacKenzie, N.~J. Balmforth, D.~M. Martinez, D.~R. Hewitt,
  Obstructed viscoplastic flow in a {H}ele-{S}haw cell, Phys. Rev. Fluids 5~(1)
  (2020) 013301.

\bibitem{OVARLEZ201368}
G.~Ovarlez, S.~Cohen-Addad, K.~Krishan, J.~Goyon, P.~Coussot, On the existence
  of a simple yield stress fluid behavior, J. Non-Newtonian Fluid Mech. 193
  (2013) 68--79.

\bibitem{Putz2009}
A.~Putz, T.~I. Burghelea, The solid--fluid transition in a yield stress shear
  thinning physical gel, Rheol. Acta 48~(6) (2009) 673--689.

\bibitem{Balmforth2014}
N.~J. Balmforth, I.~A. Frigaard, G.~Ovarlez, Yielding to stress: recent
  developments in viscoplastic fluid mechanics, Annu. Rev. Fluid Mech 46~(1)
  (2014) 121--146.

\bibitem{Coussot2014}
P.~Coussot, Yield stress fluid flows: A review of experimental data, J.
  Non-Newtonian Fluid Mech. 211 (2014) 31--49.

\bibitem{zare2021effects}
M.~Zare, M.~Daneshi, I.~A. Frigaard, Effects of non-uniform rheology on the
  motion of bubbles in a yield-stress fluid, J. Fluid Mech. 919 (2021) A25.

\bibitem{daneshi2023growth}
M.~Daneshi, I.~A. Frigaard, Growth and stability of bubbles in a yield stress
  fluid, J. Fluid Mech. 957 (2023) A16.

\bibitem{Abdali1992}
S.~S. Abdali, E.~Mitsoulis, N.~C. Markatos, Entry and exit flows of {B}ingham
  fluids, J. Rheol. 36~(2) (1992) 389--407.

\bibitem{Mitsoulis1993}
E.~Mitsoulis, S.~S. Abdali, N.~C. Markatos, Flow simulation of
  {H}erschel-{B}ulkley fluids through extrusion dies, Can. J. Chem. Eng. 71~(1)
  (1993) 147--160.

\bibitem{MITSOULIS2001173}
E.~Mitsoulis, T.~h. Zisis, Flow of {B}ingham plastics in a lid-driven square
  cavity, J. Non-Newtonian Fluid Mech. 101~(1-3) (2001) 173--180.

\bibitem{VIGNEAUX201838}
P.~Vigneaux, G.~Chambon, A.~Marly, L.~Luu, P.~Philippe, Flow of a yield-stress
  fluid over a cavity: {E}xperimental and numerical investigation of a
  viscoplastic boundary layer, J. Non-Newtonian Fluid Mech. 261 (2018) 38--49.

\bibitem{hewitt2016}
D.~R. Hewitt, M.~Daneshi, N.~J. Balmforth, D.~M. Martinez, Obstructed and
  channelized viscoplastic flow in a hele-shaw cell, J. Fluid Mech. 790 (2016)
  173--204.

\bibitem{balmforth2017}
N.~J. Balmforth, R.~V. Craster, D.~R. Hewitt, S.~Hormozi, A.~Maleki,
  Viscoplastic boundary layers, J. Fluid Mech. 813 (2017) 929--954.

\bibitem{Jay2001}
P.~Jay, A.~Magnin, J.~M. Piau, Viscoplastic fluid flow through a sudden
  axisymmetric expansion, AlChE J. 47~(10) (2001) 2155--2166.

\bibitem{Alexandrou2001}
A.~N. Alexandrou, T.~M. McGilvreay, G.~Burgos, Steady {H}erschel--{B}ulkley
  fluid flow in three-dimensional expansions, J. Non-Newtonian Fluid Mech.
  100~(1-3) (2001) 77--96.

\bibitem{DeSouzaMendes2007}
P.~R. de~Souza~Mendes, M.~F. Naccache, P.~R. Varges, F.~H. Marchesini, Flow of
  viscoplastic liquids through axisymmetric expansions--contractions, J.
  Non-Newtonian Fluid Mech. 142~(1-3) (2007) 207--217.

\bibitem{Luu2015}
L.~Luu, P.~Philippe, G.~Chambon, Experimental study of the solid-liquid
  interface in a yield-stress fluid flow upstream of a step, Phys. Rev. E
  91~(1) (2015) 013013.

\bibitem{Varges2020}
P.~R. Varges, B.~S. Fonseca, P.~R. de~Souza~Mendes, M.~F. Naccache, C.~R.
  de~Miranda, Flow of yield stress materials through annular abrupt
  expansion--contractions, Phys. Fluids 32~(8) (2020) 083101.

\bibitem{Philippe2017}
L.~Luu, P.~Philippe, G.~Chambon, Flow of a yield-stress fluid over a cavity:
  {E}xperimental study of the solid--fluid interface, J. Non-Newtonian Fluid
  Mech. 245 (2017) 25--37.

\bibitem{Tokpavi2009}
D.~L. Tokpavi, P.~Jay, A.~Magnin, L.~Jossic, Experimental study of the very
  slow flow of a yield stress fluid around a circular cylinder, J.
  Non-Newtonian Fluid Mech. 164~(1-3) (2009) 35--44.

\bibitem{Ahonguio2014}
F.~Ahonguio, L.~Jossic, j.~v. p. y.~p. Magnin, A., Influence of surface
  properties on the flow of a yield stress fluid around spheres.

\bibitem{TOKPAVI200935}
D.~L. Tokpavi, P.~Jay, A.~Magnin, L.~Jossic, Experimental study of the very
  slow flow of a yield stress fluid around a circular cylinder, J.
  Non-Newtonian Fluid Mech. 164~(1-3) (2009) 35--44.

\bibitem{JOSSIC201314}
L.~Jossic, F.~Ahonguio, A.~Magnin, Flow of a yield stress fluid perpendicular
  to a disc, J. Non-Newtonian Fluid Mech. 191 (2013) 14--24.

\bibitem{daneshi2019characterising}
M.~Daneshi, A.~Pourzahedi, D.~M. Martinez, D.~Grecov, Characterising wall-slip
  behaviour of {C}arbopol gels in a fully-developed {P}oiseuille flow, J.
  Non-Newtonian Fluid Mech. 269 (2019) 65--72.

\bibitem{cheddadi}
I.~Cheddadi, P.~Saramito, B.~Dollet, C.~Raufaste, F.~Graner, Understanding and
  predicting viscous, elastic, plastic flows, Eur. Phys. J. E 34~(1) (2011)
  1--15.

\bibitem{Fraggedakis}
D.~Fraggedakis, Y.~Dimakopoulos, J.~Tsamopoulos, Yielding the yield-stress
  analysis: a study focused on the effects of elasticity on the settling of a
  single spherical particle in simple yield-stress fluids, Soft matter 12~(24)
  (2016) 5378--5401.

\bibitem{izbassarov2018computational}
D.~Izbassarov, M.~E. Rosti, M.~N. Ardekani, M.~Sarabian, S.~Hormozi, L.~Brandt,
  O.~Tammisola, Computational modeling of multiphase viscoelastic and
  elastoviscoplastic flows, Int. J. Numer. Meth. Fluids 88~(12) (2018)
  521--543.

\bibitem{chaparian2019porous}
E.~Chaparian, D.~Izbassarov, F.~De~Vita, L.~Brandt, O.~Tammisola, Yield-stress
  fluids in porous media: a comparison of viscoplastic and elastoviscoplastic
  flows, Meccanica 55 (2020) 331--342.

\bibitem{chaparian2019adaptive}
E.~Chaparian, O.~Tammisola, An adaptive finite element method for
  elastoviscoplastic fluid flows, J. Non-Newtonian Fluid Mech. 271 (2019)
  104148.

\bibitem{SARAMITO2007}
P.~Saramito, A new constitutive equation for elastoviscoplastic fluid flows, J.
  Non-Newtonian Fluid Mech. 145~(1) (2007) 1--14.

\bibitem{saramito2009new}
P.~Saramito, A new elastoviscoplastic model based on the {H}erschel--{B}ulkley
  viscoplastic model, J. Non-Newtonian Fluid Mech. 158~(1-3) (2009) 154--161.

\bibitem{bingham1922}
E.~C. Bingham, Fluidity and plasticity, Vol.~2, McGraw-Hill, 1922.

\bibitem{oldroyd1950}
J.~G. Oldroyd, On the formulation of rheological equations of state, Proc. R.
  Soc. London, Ser. A 200~(1063) (1950) 523--541.

\bibitem{COUSSOT201431}
P.~Coussot, Yield stress fluid flows: {A} review of experimental data, J.
  Non-Newtonian Fluid Mech. 211 (2014) 31--49.

\bibitem{Lectures_Ovarlez}
G.~Ovarlez, Rheometry of visco-plastic fluids, Lectures on Visco-Plastic Fluid
  Mechanics (2019) 127--163.

\bibitem{Carbopol_divoux}
R.~Radhakrishnan, T.~Divoux, S.~Manneville, S.~M. Fielding, Understanding
  rheological hysteresis in soft glassy materials, Soft Matter 13~(9) (2017)
  1834--1852.

\bibitem{hormozi2011}
S.~Hormozi, D.~Martinez, I.~Frigaard, Stable core-annular flows of viscoelastic
  fluids using the visco-plastic lubrication technique, J. Fluid Mech.
  166~(23-24) (2011) 1356--1368.

\bibitem{OCT1}
R.~A. Leitgeb, R.~M. Werkmeister, C.~Blatter, L.~Schmetterer, Doppler optical
  coherence tomography, Progress in retinal and eye research 41 (2014) 26--43.

\bibitem{ConfocalI}
R.~H. Webb, Confocal optical microscopy, Reports on progress in physics 59~(3)
  (1996) 427.

\bibitem{ConfocalII}
R.~Lima, S.~Wada, K.-i. Tsubota, T.~Yamaguchi, Confocal micro-piv measurements
  of three-dimensional profiles of cell suspension flow in a square
  microchannel, Measurement Science and Technology 17~(4) (2006) 797.

\bibitem{daneshi2020thesis}
M.~Daneshi, Characterising the non-ideal behaviour of a carbopol gel flowing in
  thin conduits, Ph.D. thesis, University of British Columbia (2020).

\bibitem{PIVmethod}
F.~Scarano, M.~L. Riethmuller, Advances in iterative multigrid piv image
  processing, Experiments in fluids 29~(Suppl 1) (2000) S051--S060.

\bibitem{drost2013hele}
S.~Drost, J.~Westerweel, Hele-shaw rheometry, J. Rheol. 57~(6) (2013)
  1787--1801.

\bibitem{kim2000instabilities}
J.~Kim, A.~{\"O}ztekin, S.~Neti, Instabilities in viscoelastic flow past a
  square cavity, J. Non-Newtonian Fluid Mech. 90~(2-3) (2000) 261--281.

\bibitem{yamamoto2003}
T.~Yamamoto, M.~Ishiyama, M.~Nakajima, K.~Nakamura, N.~Mori, Three-dimensional
  viscoelastic flows through a rectangular channel with a cavity, J.
  Non-Newtonian Fluid Mech. 114~(1) (2003) 13--31.

\bibitem{roquet2003adaptive}
N.~Roquet, P.~Saramito, An adaptive finite element method for {B}ingham fluid
  flows around a cylinder, Comput. Meth. Appl. Mech. Eng. 192~(31) (2003)
  3317--3341.

\bibitem{MR3043640}
F.~Hecht, New development in {F}ree{F}em++, J. Numer. Math. 20~(3) (2012)
  251--265.

\bibitem{chaparian2017yield}
E.~Chaparian, I.~A. Frigaard, Yield limit analysis of particle motion in a
  yield-stress fluid, J. Fluid Mech. 819 (2017) 311--351.

\bibitem{iglesias2020computing}
J.~A. Iglesias, G.~Mercier, E.~Chaparian, I.~A. Frigaard, Computing the yield
  limit in three-dimensional flows of a yield stress fluid about a settling
  particle, J. Non-Newtonian Fluid Mech. 284 (2020) 104374.

\bibitem{chaparian2020sliding}
E.~Chaparian, O.~Tammisola, Sliding flows of yield-stress fluids, J. Fluid
  Mech. 911 (2021) A17.

\bibitem{chaparian2020particle}
E.~Chaparian, M.~N. Ardekani, L.~Brandt, O.~Tammisola, Particle migration in
  channel flow of an elastoviscoplastic fluid, J. Non-Newtonian Fluid Mech. 284
  (2020) 104376.

\bibitem{chaparian2017cloaking}
E.~Chaparian, I.~A. Frigaard, Cloaking: Particles in a yield-stress fluid, J.
  Fluid Mech. 243 (2017) 47--55.

\end{thebibliography}


%\begin{thebibliography}{00}



%\end{thebibliography}
\end{document}